\documentclass[journal]{IEEEtran}
\usepackage{amsmath,amsfonts}
\usepackage{algorithmic}
\usepackage{algorithm}
\usepackage{array}
\usepackage{booktabs}
\usepackage{threeparttable}
\usepackage[caption=false,textfont={rm}]{subfig}
\usepackage{textcomp}
\usepackage{stfloats}
\usepackage{url}
\usepackage{verbatim}
\usepackage{graphicx}
\usepackage{cite}
\usepackage{gensymb}
\usepackage{multirow}
\usepackage{makecell}
\usepackage{float}
\usepackage{color}
\usepackage{bm}
\usepackage{color}
\usepackage{upgreek}
\usepackage{tabularx}

\begin{document}

\title{Comparison and Analysis of Cognitive Load under 2D/3D Visual Stimuli}

\author{Yu~Liu, 
        Chen~Song, 
        Yunpeng~Yin, 
        Herui~Shi,
        Jinglin~Sun,      
        Han~Wang,
        Minpeng~Xu
\thanks{  
\textit{Corresponding author: Yu Liu, liuyu@tju.edu.cn.}} }

\markboth{}%
{Shell \MakeLowercase{\textit{et al.}}: A Sample Article Using IEEEtran.cls for IEEE Journals}


\maketitle

\begin{abstract}

With the increasing prevalence of 3D videos, investigating the differences of viewing experiences between 2D and 3D videos has become an important issue. In this study, we explored the cognitive load induced by 2D and 3D video stimuli under various cognitive tasks utilizing electroencephalogram (EEG) data. We also introduced the Cognitive Load Index (CLI), a metric which combines $\theta$ and $\alpha$ oscillations to evaluate the cognitive differences. Four video stimuli, each associated with typical cognitive tasks were adopted in our experiments. Subjects were exposed to both 2D and 3D video stimuli, and the corresponding EEG data were recorded. Then, we analyzed the power within the 0.5-45 Hz frequency of EEG data, and CLI was utilized to evaluate the brain activity of different subjects.
According to our experiments and analysis, videos that involve simple observational tasks ($P$ \textless 0.05) consistently induced a higher cognitive load in subjects when they were viewing 3D videos. However, for videos that involve calculation tasks ($P$ \textgreater 0.05), the differences in cognitive load induced by 2D and 3D video were not obvious. Thus, we concluded that 3D videos could generally induce a higher cognitive load, but the extent of the differences also depended on the contents of the video stimuli and the viewing purpose.

\end{abstract}

\begin{IEEEkeywords}
EEG, Cognitive load, $\alpha$ oscillation, $\theta$ oscillation, 3D video
\end{IEEEkeywords}

\section{Introduction}
\label{sect_intro}
\IEEEPARstart{C}{OGNITION} is a fundamental psychological function in humans that encompasses the processes of acquiring, interpreting, and utilizing information. The cognitive load, or the amount of mental effort required to process input information, always increases with the complexity of cognitive tasks. The cognitive load is significant in the design of human-machine interacting systems\cite{GALY2012269,aygun2022investigating}.
Some questionnaire measuring methods for cognitive load had been proposed \cite{krieglstein2023development,paas1994measurement,paas2003cognitive,paas2016cognitive}.
Besides, physiological indicators-based methods for assessing cognition utilizing parameters such as heart rate\cite{ma2024determining}, eye movements\cite{tong2023measuring,liu2022assessing}, hormone levels\cite{wilsonft}, functional magnetic resonance imaging (fMRI) and electroencephalography (EEG)\cite{klimesch1999eeg,anderson2008translating} were also proposed.
However, physiological methods for measuring cognitive load are limited in several aspects. 
For instance, heart rate variations and hormone level may not sensitively reflect transient changes in cognitive load. Similarly, the responsiveness of pupil dilation to cognitive load may decrease with the age of the subjects\cite{hartley2022equivalent}.
Comparing with these methods, fMRI and EEG could provide sufficient information for brain activities. Further considering the testing convenience, the EEG test is more widely adopted as general method for evaluating cognitive load under different stimuli.

3D videos have become more and more popular, and investigating the differences in the viewing experiences between 2D and 3D videos also has become an important issue.
Some studies had been carried out to investigate the differences between subjects under 2D and 3D video stimuli. For instance, the study conducted by Slobounov et al. \cite{slobounov2015modulation} had revealed that within 3D video stimuli, the frontal lobe of the brain exhibited a higher average $\theta$ oscillation power compared to those in 2D video stimuli.
Similarly, Alex Dan's study \cite{pmid27592084} had unveiled statistically significant variations in $\beta$ oscillation activity among subjects viewing 3D videos compared to their counterparts viewing 2D videos. Notably, these distinctions appeared to be contingent on the nature of the video content, with content characterized by higher velocity and more dynamic elements tending to elicit more significant $\beta$ oscillation activity. Moreover, the study conducted by Chunxiao Chen \cite{chen2013eeg} had contributed additional insights on this subject. Chen's study had demonstrated that subjects exposed to 3D videos underwent more prominent alterations in specific brain parameters compared to those viewing 2D videos.
These alterations manifested as a notable decrease in power within the $\alpha$ and $\beta$ oscillations,  while the $\theta$ oscillation activity generally remained relatively stable. 

Existing studies often neglect the systematic categorization of video stimuli. In this paper, we utilized EEG to investigate differences in cognitive load when subjects viewing 2D and 3D videos, which were classified into four types according to the contents. Besides, we also introduced a specialized Cognitive Load Index (CLI) to evaluate the variations of EEG patterns when subjects were viewing different videos stimuli.

In summary, the main contribution of this paper can be summarized as follows.

\begin{itemize}
\item In the experiment setup, four video stimuli featuring distinct characteristics and typical cognitive tasks were adopted to induce cognitive load. Based on the characteristics and cognitive tasks, the video stimuli were carefully classified to explore the factors influencing cognitive load.

\item A credible experiment was designed to explore the differences in cognitive load under 2D and 3D stimuli. In this paper, cognitive load was evaluated by CLI, and the relationship between cognitive load and 2D/3D video stimuli was summarized.

\item In this paper, we conducted statistical analysis on the CLI, as well as the power of various oscillations at Fz and Pz electrodes. The results revealed the factors affecting cognitive load and the features of different video stimuli reflected in EEG signals.

\end{itemize}

The remainder of this paper is organized as follows. Section II briefly reviews studies associated with Cognitive Load Index (CLI) and EEG. Section III provides a detailed explanation of the experimental process, data processing, and statistical analysis methods. The experimental data and statistical analysis results under different video stimuli are presented in Section IV, followed by a discussion in Section V. Finally, the conclusion is presented in Section VI.

\section{EEG and Cognitive Load Index}

Among all EEG oscillations, $\alpha$ and $\theta$ oscillations are particularly pertinent in cognitive processes \cite{basar1999brain,klimesch2005functional,penfield1954epilepsy}, reflecting subject's cognitive states during information processing.
$\alpha$ oscillation is associated with spatial awareness, working memory, and attention \cite{kober2012cortical,klimesch1997eeg,kweon2017brain}. While, $\theta$ oscillation is associated with episodic memory \cite{buzsaki1994oscillatory,miller1991cortico}. Klimesch \cite{klimesch1999eeg} proposed that $\theta$ oscillation was involved in the reception of new information, suggesting that when subjects encode novel information, $\theta$ oscillation power always increase significantly.
Miller conducted a study involving semantic consistency and situational recognition tasks \cite{miller1991cortico}, where EEG data was analyzed utilizing event-related desynchronization to gauge the consistent correlation variations in $\theta$ and $\alpha$ oscillation power. The experimental results indicated that situational information encoding correlated with $\theta$ oscillation, while semantic memory processes were associated with $\alpha$ oscillation.

As previously mentioned, $\alpha$ and $\theta$ oscillations play crucial roles in cognition. Hence, investigating EEG alterations under 2D and 3D video stimuli requires the analysis of $\alpha$ and $\theta$ oscillations \cite{dan2017eeg,kweon2017brain,chen2013eeg,manshouri2016classification}. However, the conclusion of these studies are not consistent.
In Kweon's study \cite{kweon2017brain}, subjects viewing 2D videos exhibited higher $\alpha$ oscillation power compared to those viewing 3D videos. Conversely, in Dan's study \cite{dan2017eeg}, subjects demonstrated higher $\alpha$ oscillation power when engaging in complex tasks during 3D videos viewing, while $\alpha$ oscillation power remained similar for less complex tasks in both 2D and 3D videos. Besides, Klimesch \cite{klimesch1999eeg} declared that EEG patterns, such as $\alpha$ oscillation, exhibited significant individual variability, and  utilizing the average value as the basis for EEG data comparison might obscure the characteristics of small samples.

Cognitive Load Index (CLI) can be defined as the ratio of the absolute power of $\theta$ oscillation to $\alpha$ oscillation, so that the  influence of these two oscillations on cognitive load can be combined. Holm et al. \cite{holm2009estimating} suggested, in the comparison of CLI, that $\theta$ oscillation should be quantified by the absolute power of $\theta$ oscillation in the subject's frontal lobe (Fz), while $\alpha$ oscillation should be calculated by the absolute power of $\alpha$ oscillation in the subject's parietal lobe (Pz). They declared that this ratio was an effective indicator of cognitive load. Besides, a higher CLI suggested more efficient brain activity.
In this paper, the CLI is defined as:
$$CLI=\frac {\theta_{Fz}} {\alpha_{Pz}}$$

\section{Evaluating Experiments Setup and Analysing Methods}
\label{sect_method}

\subsection{Subjects}
\label{sect_method:subjects}
30 subjects (17 males and 13 females) free from visual impairments attended our experiments.
The age of the subjects varied between 21 and 25, with a mean age of 22.82. Prior to the formal experiment, each subject viewed an unrelated 3D video utilizing our 3D playback device to ensure they could effectively perceive 3D visuals.
Furthermore, subjects were instructed to prioritize quality rest and abstain from alcohol and caffeine.

\begin{figure}[!t]
    \centering
    \includegraphics[width=1\linewidth]{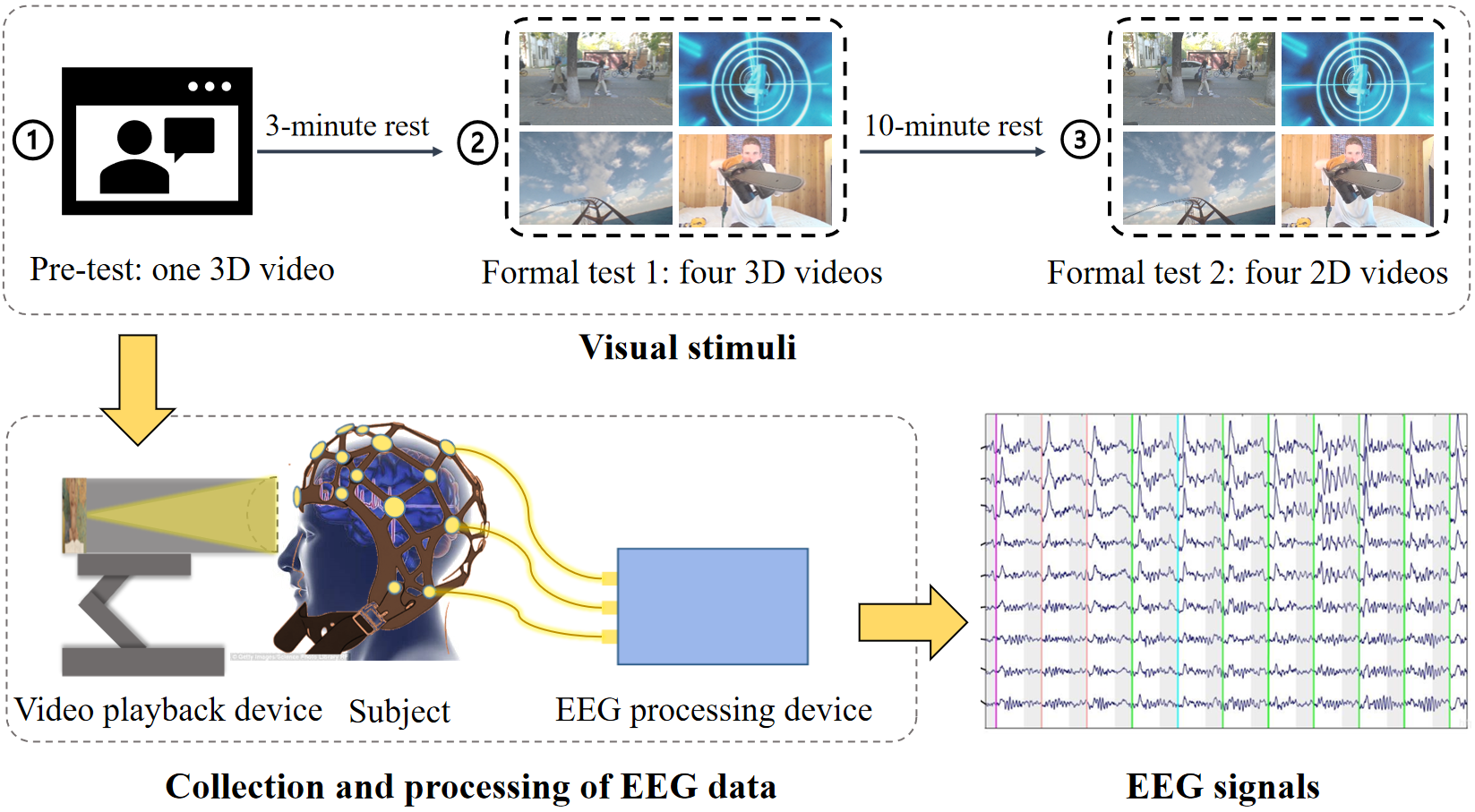}
    \caption{Illustration of the overall experimental process. The subjects wear a specialized EEG cap to record EEG data, utilizing a customized video play devices which can provide 2D/3D stimuli. The testing procedure comprises a pre-test video for warm-up, followed by two formal test sessions: the first consists of four 3D videos and the second consists of four 2D videos showing the same contents as the 3D videos.
    }
    \label{fig_pro}
\end{figure}

\subsection{Video stimuli}
\label{sect_method:Video stimuli}
\begin{figure}[!t]
    \centering
    \subfloat[Video stimuli (a)] {\includegraphics[width=0.235\textwidth]{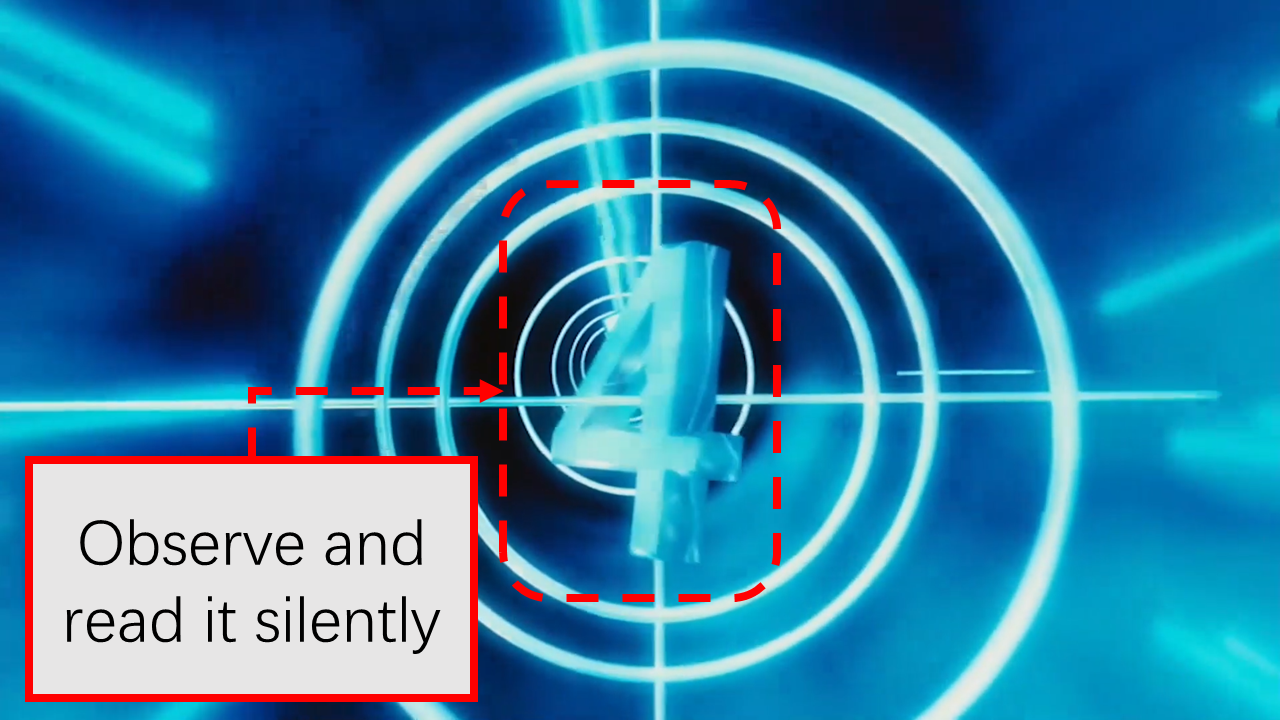}}
    \hfill
    \subfloat[Video stimuli (b)] {\includegraphics[width=0.235\textwidth]{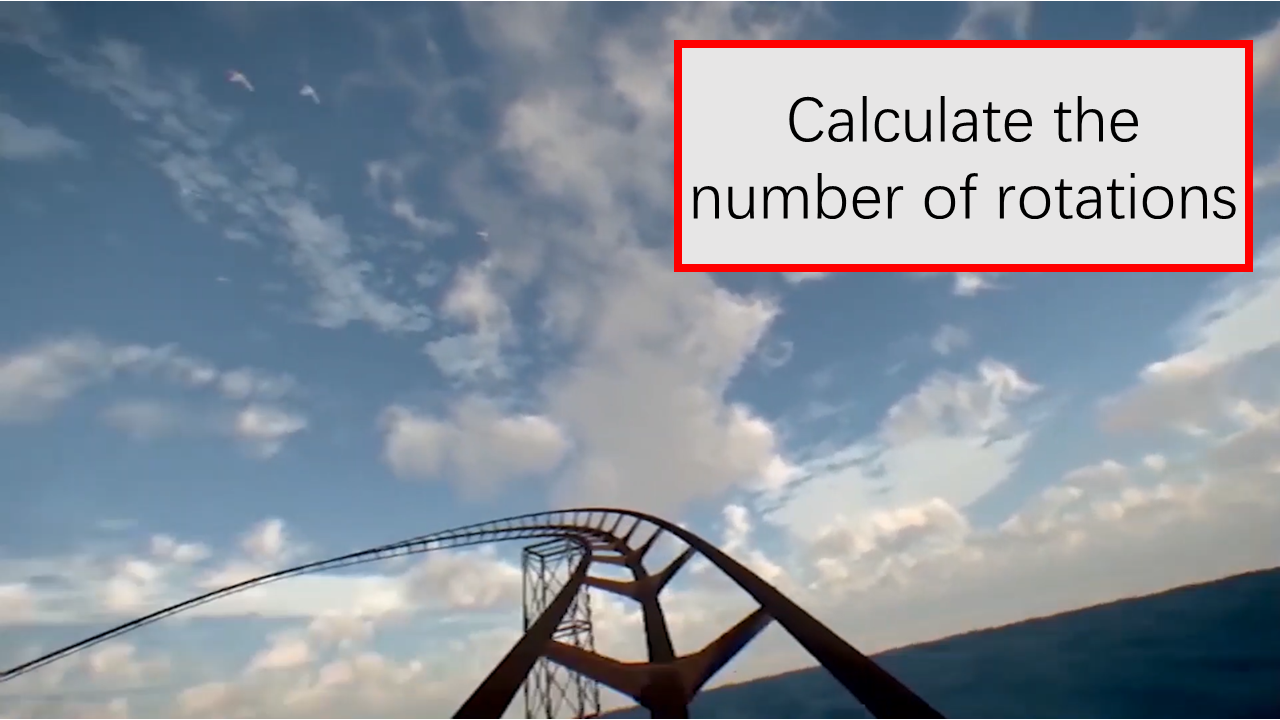}}
    \newline
    \subfloat[Video stimuli (c)] {\includegraphics[width=0.235\textwidth]{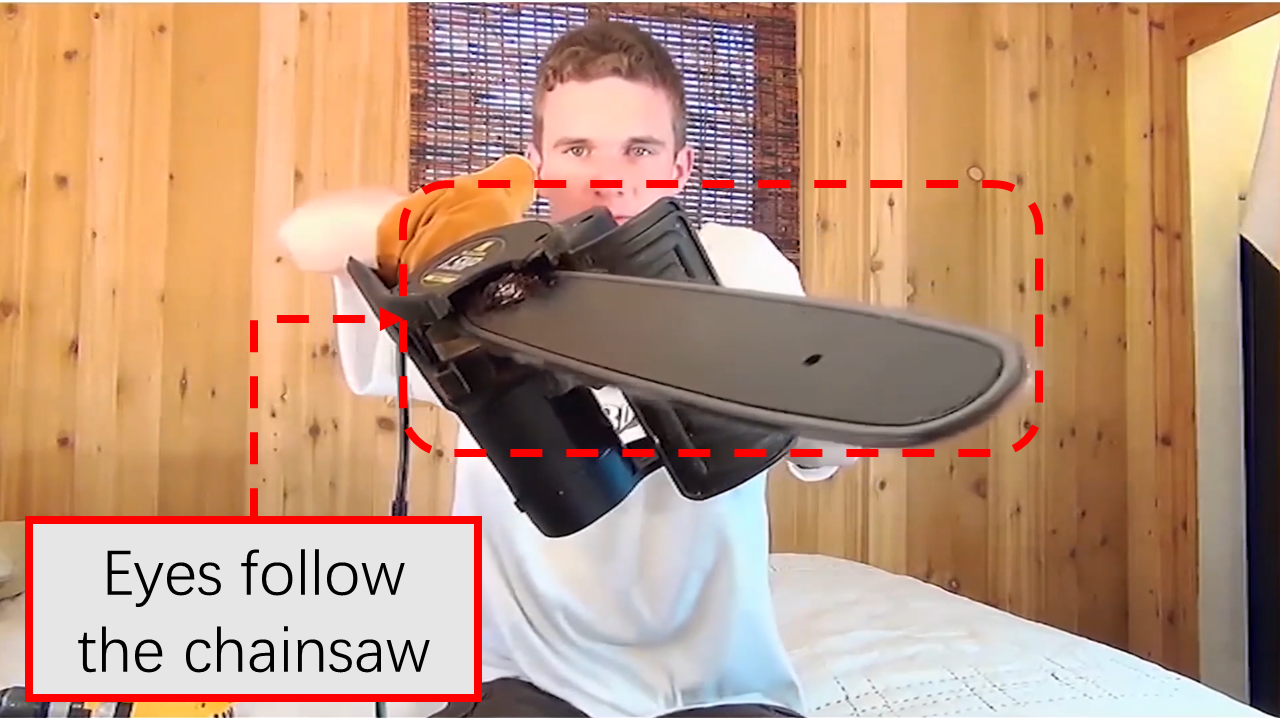}}
    \hfill
    \subfloat[Video stimuli (d)] {\includegraphics[width=0.235\textwidth]{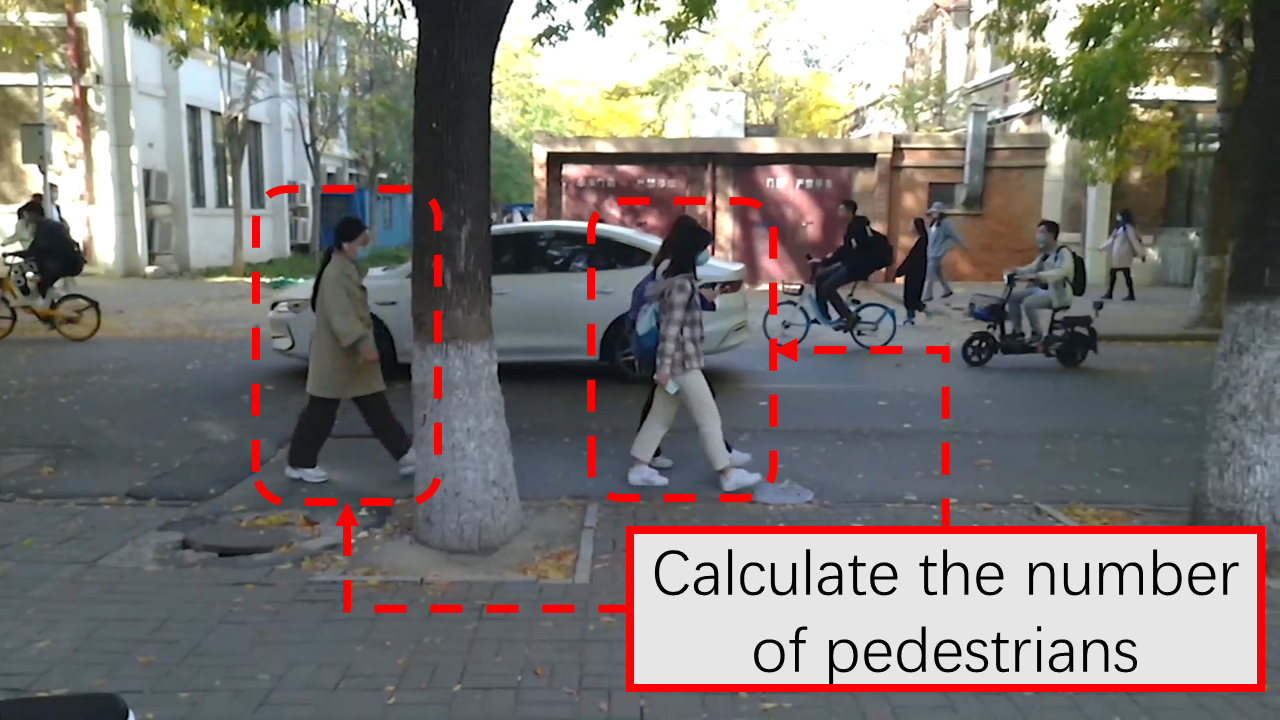}}
    \caption{Video stimuli and corresponding tasks.}
    \label{fig:distr}
\end{figure}

Numerous studies investigating the disparities in EEG patterns between subjects viewing 2D and 3D videos utilized video classification methods \cite{dan2017eeg,kweon2017brain}.
On one hand, viewers typically engage with videos with a specific purpose or task in mind. On the other hand, the level of content dynamism within the videos can vary significantly.
Hence, videos can be categorized based on two criteria: the intended purpose of viewing and the content characteristics.
In terms of viewing purpose, videos can be categorized into those intended for simple observational tasks and those aimed at simple calculation tasks.
Whereas, considering the contents, videos can be categorized into two categories: those featuring rapid and dynamic changes in content and those characterized by a more serene and steady pace. These ``dynamic changes'' may involve alterations in the displayed objects or shifts in viewing perspective.
The video stimuli utilized in this experiment were categorized into four sections, each with both 2D and 3D stimuli.
By defining video stimuli (a) - (d) corresponding with four distinct videos respectively, the introduction and classification of experimental video stimuli were shown in Table \ref{tab_class}.
Screen shot of the stimuli were shown in Fig.~\ref{fig:distr}.

\begin{table*}[!t]
    \caption{Four video stimuli}
    \centering
    \begin{tabular}
    {ccccccc}
        \toprule
        \textbf{Video stimuli} & \textbf{Duration}& \textbf{Content}& \textbf{Characteristics}& \textbf{Viewing tasks}& \textbf{Category I}& \textbf{Category II}\\
        \midrule
        \makecell*[c]{Video stimuli (a)}
        & \makecell*[c]{30 sec}
        & \makecell*[l]{This video contains\\ huge Arabic numerals.}
        & \makecell*[l]{The number in the video will\\ change from 9 to 1.}
        & \makecell*[l]{Subjects were instructed to\\ mentally count down follow-\\ing the numerical changes.}
        & \makecell*[l]{Video with simple\\ observation task}
        & \makecell*[c]{Tranquil video}\\
        \midrule
         \makecell*[c]{Video stimuli (b)}
         & \makecell*[c]{30 sec}
         & \makecell*[l]{This video features\\ a virtual roller coa-\\ ster ride.}
         & \makecell*[l]{The viewing perspective\\ changes frequently. At times, \\the rotation may be 180\\ degrees, and at other times, \\360 degrees. In addition,\\ the perspective of the video \\periodically rotates several\\ 360 degrees continuously\\ and quickly.}
         & \makecell*[l]{Subjects were asked to calcu-\\late the total number of 360\\ degrees of rotation, consid-\\ering every two 180 degrees\\ rotations as one 360 degrees\\ rotation, regardless of their\\ continuity.}
         & \makecell*[l]{Video with simple\\ calculation task}
         & \makecell*[l]{High-dynamic\\ video}\\
        \midrule
        \makecell*[c]{Video stimuli (c)}
        & \makecell*[c]{30 sec}
        & \makecell*[l]{This video shows a\\ man operating an c-\\ hainsaw.}
        & \makecell*[l]{The chainsaw moves slowly\\ during operation.}
        & \makecell*[l]{Subjects were asked to keep\\ an eye on the chainsaw at all\\ times.}
        & \makecell*[l]{Video with simple\\ observation task}
        & \makecell*[c]{Tranquil video}\\
        \midrule
         \makecell*[c]{Video stimuli (d)}
         & \makecell*[c]{30 sec}
         & \makecell*[l]{This video shows b-\\ ustling street crowd-\\ ed with pedestrians.}
         & \makecell*[l]{The pedestrians and vehi-\\cles in the video change fre-\\quently.}
         & \makecell*[l]{Subjects were asked to count\\ the number of people moving\\ from left to right. Any car\\ appearing in the video should\\ be counted as two people.}
         & \makecell*[l]{Video with simple\\ calculation task}
         & \makecell*[l]{High-dynamic\\ video}\\
        \bottomrule
    \end{tabular}
\label{tab_class}
\end{table*}

\subsection{Video playback device}
\label{sect_method:Video playback device}
The video playback device we utilized is a desktop with dimensions of 412~$\rm mm$ in length, 270~$\rm mm$ in width, and 105~$\rm mm$ in thickness. 
The device can provide excellent 3D viewing experiences \cite{9964289}.
Like most VR devices, this video playback device achieves 3D visuals by providing left/right images separately to the left/right eye of subjects. When subjects use the device to view videos, their visual field is constrained within the screen, so that their visual attention is only associated with the video stimuli.

\subsection{Procedure}
\label{sect_method:Procedures}
\begin{figure}[!t]
    \centering
    \includegraphics[width=1\linewidth]{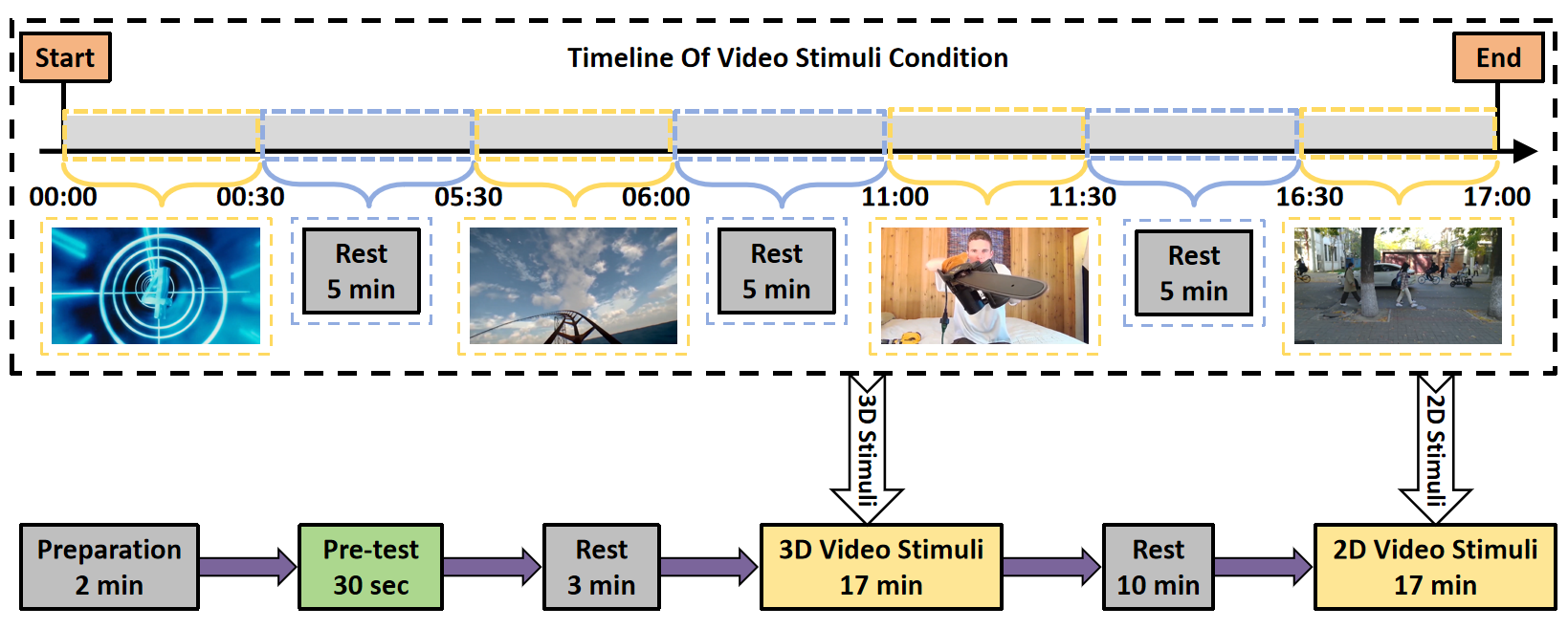}
    \caption{Timeline of the experiment. The figure shows the chronological order in which subjects view the video stimuli.
    }
    \label{fig_time}
\end{figure}

\begin{figure}[!t]
    \centering
    \includegraphics[width=1\linewidth]{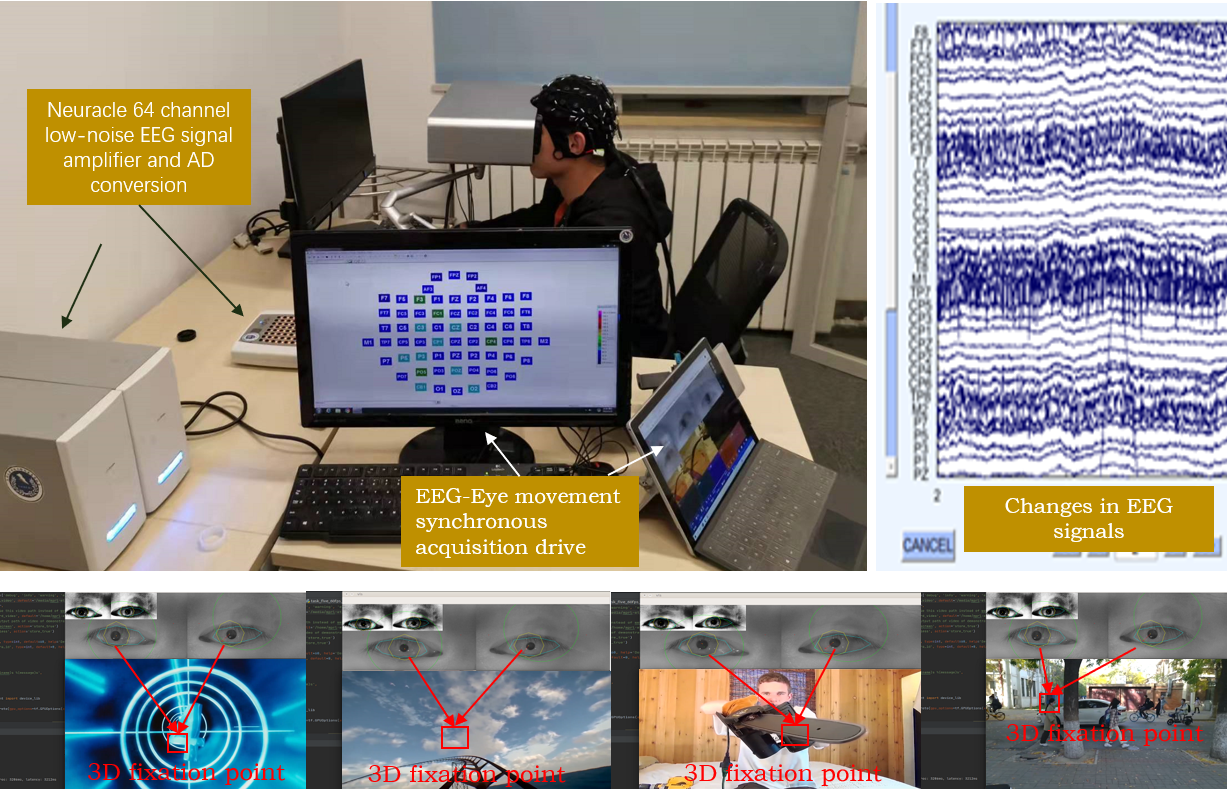}
    \caption{Experimental environments, 3D eye movement analysis and EEG data recording.
    }
    \label{fig_drive}
\end{figure}

Before the experiment, subjects will be provided with instructions outlining the specific tasks involved in the experiment. The experiment will begin once subjects confirm their understanding of these tasks.
The experimental procedure is depicted in Fig.~\ref{fig_pro}. Subjects will be seated in front of the video playback device as Fig.~\ref{fig_drive}, with an EEG cap fitted and appropriately calibrated on their heads.
The time flow of viewing video stimuli is shown in Fig.~\ref{fig_time}.

\begin{itemize}
\item $Step1$: Preparation, each subject will be seated quietly with their eyes closed for two minutes. This step aims to induce a relaxation state in the subject.

\item $Step2$: Pre-test, each subject will view an unrelated 3D video to ensure that he/she can effectively perceive 3D visuals and complete the 3D calibration of video playback device.

\item $Step3$: Rest for 3 minutes, during which the subject require to recall the experimental tasks and maintain a relaxation state.

\item $Step4$: View the four 3D video stimuli (a)$-$(d) sequentially in 17 minutes. There will be a 5-minute rest between two video stimuli to decrease the impact of various video stimuli on EEG signals. During this period, subjects are required to maintain a relaxation state and will be prompted with the specific tasks for the upcoming video stimuli to ensure their recall of the experimental requirements.

\item $Step5$: Rest for 10 minutes to maintain the subjects in a relaxation state and mitigate the impact of 3D video stimuli on EEG signals during 2D video stimuli.

\item $Step6$: View the four 2D video stimuli (a)$-$(d) sequentially in 17 minutes using similar procedure as $Step4$.
\end{itemize}

The experimental environments and one of the subjects who attended the experiment are shown in Fig.~\ref{fig_drive}. During the experiment, the EEG data were recorded with an EEG recorder for the following analysis of brain activity induced by 2D/3D video stimuli. 


\subsection{EEG recording and pre-processing}
\label{sect_method:EEG recording and pre-processing}
According to the international 10-10 system, we utilized a 64-channel EEG recorder (Neuracle Technology) to capture and record EEG signals at a sampling rate of 1000 Hz as subjects viewing the video stimuli. 
In this experiment, we collected 240 (30$\times$4$\times$2) sets of EEG signals from 30 subjects. Since this experiment aims to analyze cognitive load, we primarily recorded the Fz and Pz electrodes. M1 and M2 electrodes were utilized for re-reference in subsequent preprocessing.
Data preprocessing was conducted utilizing the EEGLAB toolbox  \cite{delorme2004eeglab}, involving re-reference, filtering, baseline removal, and artifact removal. We also applied a 0.5-45 Hz band-pass filter for the filtering stage. Then, independent component analysis (ICA) was utilized to eliminate artifacts, primarily eye movement and muscle artifacts.
Ultimately, the mean power of the $\delta$ (0.5-4 Hz), $\alpha$ (4-8 Hz), $\theta$ (8-12 Hz), $\beta$ (12-30 Hz), and $\gamma$ (30-45 Hz) oscillations in the EEG signals were calculated utilizing Fourier transform.

\subsection{Statistical analysis}
\label{sect_method:Statistical Analysis}
In this experiment, we calculated the power of various oscillations and obtained the CLI value under various video stimuli. Through statistical analysis of the power differences among these oscillations corresponding to various categories of videos under both 2D and 3D stimuli, we identified the unique characteristics of various video stimuli in EEG signals. Then, we utilized box plots to statistically analyze the power of $\alpha$ and $\theta$ oscillations, as well as the CLI value. This approach can clearly observe the extreme values and overall distribution differences under various video stimuli. Moreover, we conducted T-test to analyze the differences in CLI among videos (a)$-$(d) under both 2D and 3D stimuli, and explored the distribution characteristics of the data through mean and variance analysis.

\section{Results}
\label{sect_Results}

\subsection{Brain activity differences}
\begin{figure}[!t]
    \centering
    \subfloat[Video stimuli (a)] {\includegraphics[width=0.235\textwidth]{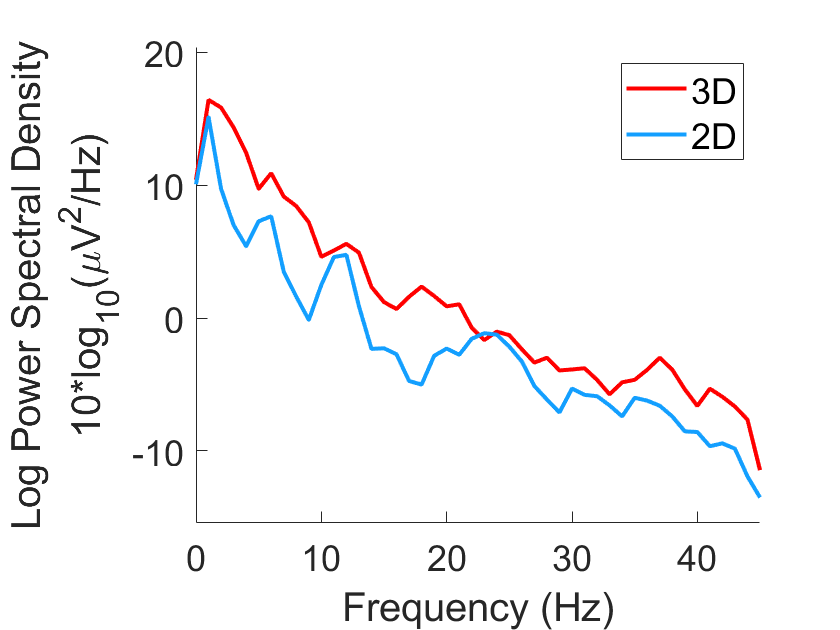}}
    \hfill
    \subfloat[Video stimuli (b)] {\includegraphics[width=0.235\textwidth]{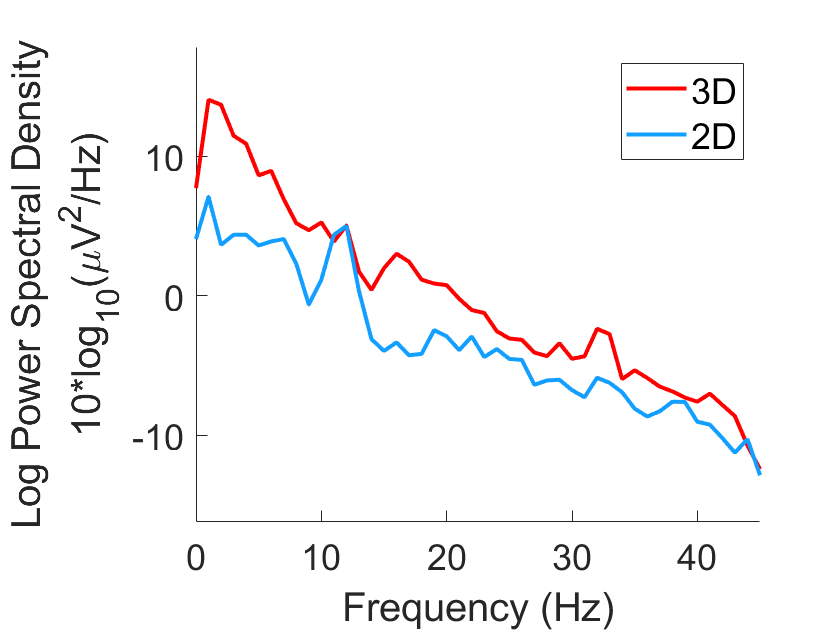}}
    \newline
    \subfloat[Video stimuli (c)] {\includegraphics[width=0.235\textwidth]{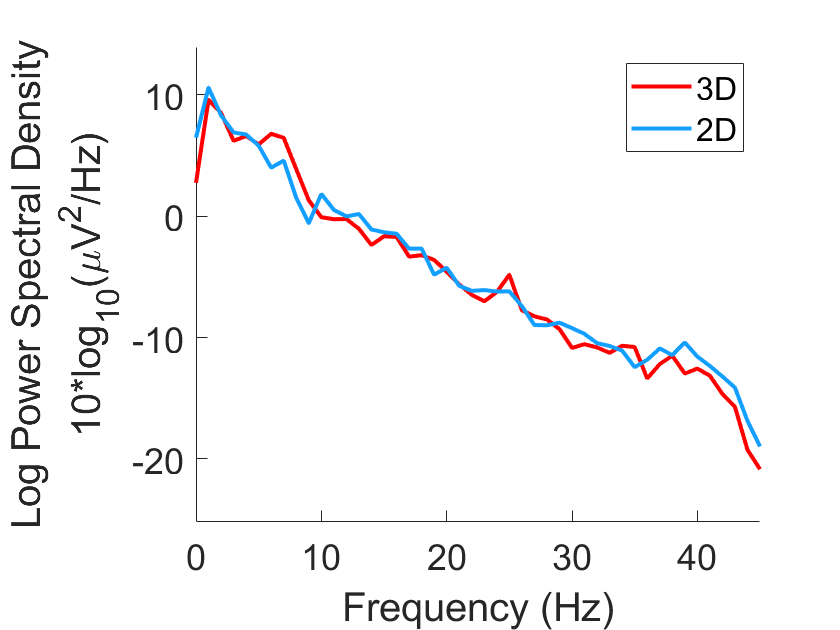}}
    \hfill
    \subfloat[Video stimuli (d)] {\includegraphics[width=0.235\textwidth]{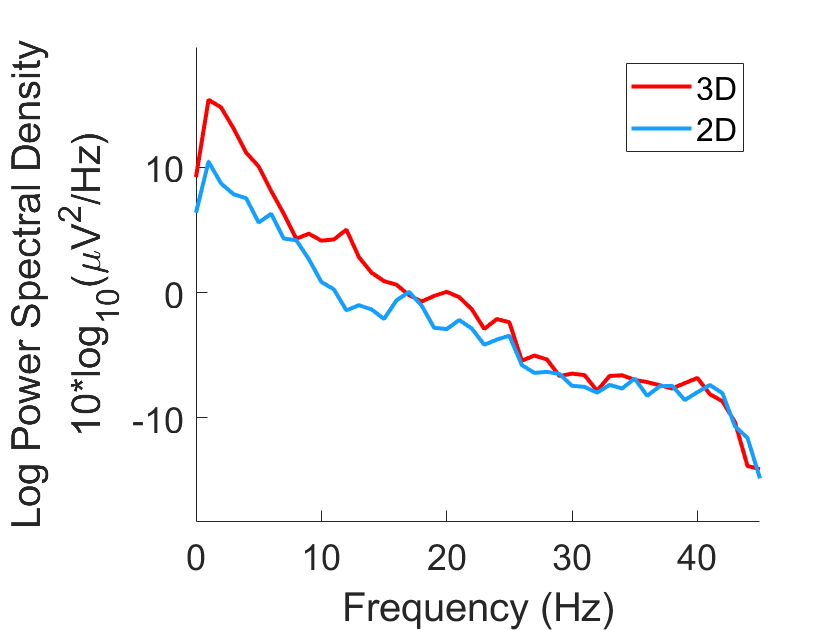}}
    \caption{The power spectral density of Fz under 2D and 3D stimuli.}
    \label{fig_fz}
\end{figure}

We analyzed the EEG data utilizing Fast Fourier Transform (FFT) to derive the power spectral density during various video stimuli. Given the strong association between cognition and the Fz and Pz electrodes, our analysis was confined to exploring the relationships between different oscillations and video stimuli at these two electrode sites.
The data of one subject is shown in Fig.~\ref{fig_fz}. The comprehensive analysis for data of the Fz electrode: 

\begin{itemize}

\item $Video$ $stimuli$ $(a)$: About 76\% of subjects exhibited higher $\theta$ oscillation power, and 69\% showed higher $\alpha$ oscillation power under 3D stimuli.

\item $Video$ $stimuli$ $(b)$: Approximately 59\% of subjects experienced higher $\beta$ oscillation power under 3D stimuli.

\item $Video$ $stimuli$ $(c)$: Around 72\% of subjects demonstrated higher $\theta$ oscillation power, and 62\% had higher $\alpha$ oscillation power under 3D stimuli. Besides, roughly 66\% of subjects showed higher $\delta$ oscillation power under 3D stimuli.

\item $Video$ $stimuli$ $(d)$: About 62\% subjects had higher $\beta$ oscillation power and 69\% showed higher $\delta$ oscillation power under 3D stimuli.

\end{itemize}

The statistical results indicated that 3D video viewing tended to enhance $\alpha$ and $\theta$ activity, and $\beta$ oscillation power increased during calculation tasks. Furthermore, for more tranquil video content, the $\delta$ oscillation power was higher under 3D stimuli.

\begin{figure}[!t]
    \centering
    \subfloat[Video stimuli (a)] {\includegraphics[width=0.235\textwidth]{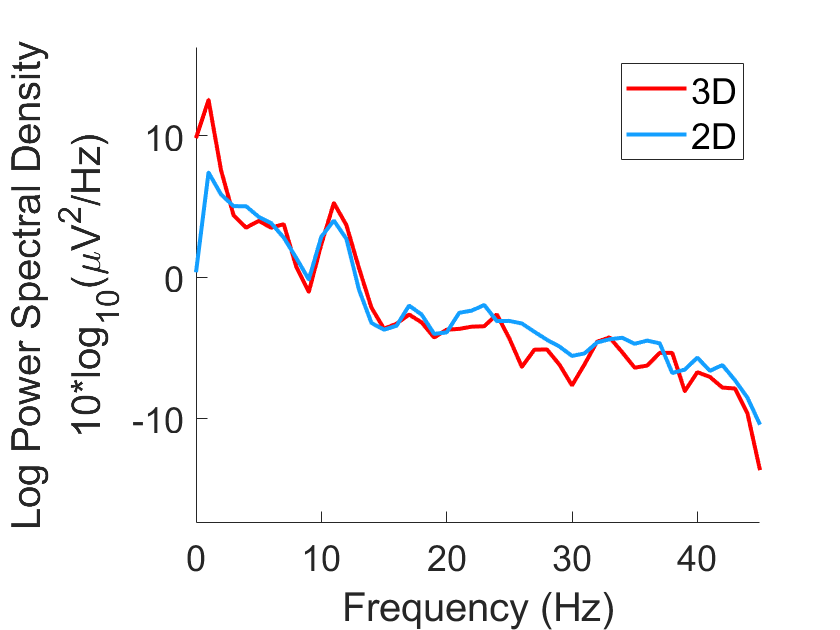}}
    \hfill
    \subfloat[Video stimuli (b)] {\includegraphics[width=0.235\textwidth]{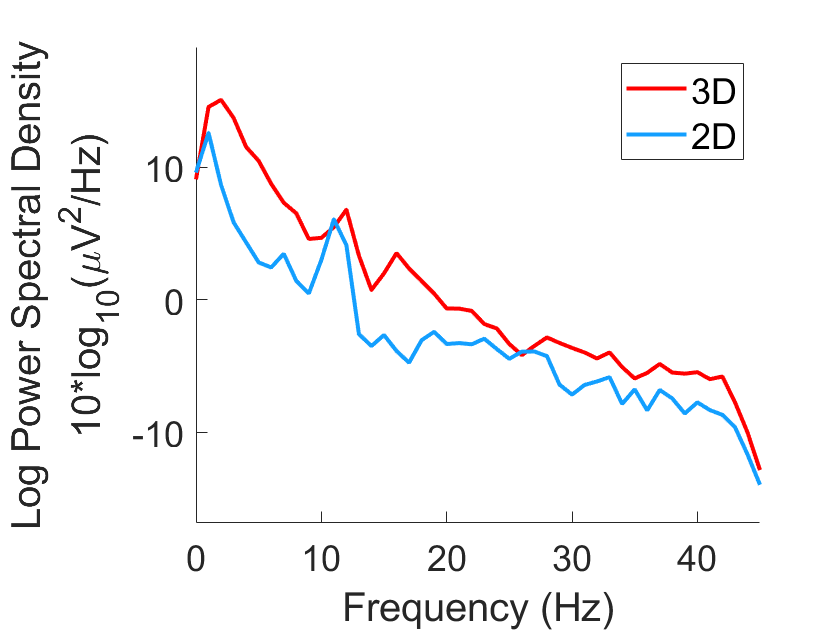}}
    \newline
    \subfloat[Video stimuli (c)] {\includegraphics[width=0.235\textwidth]{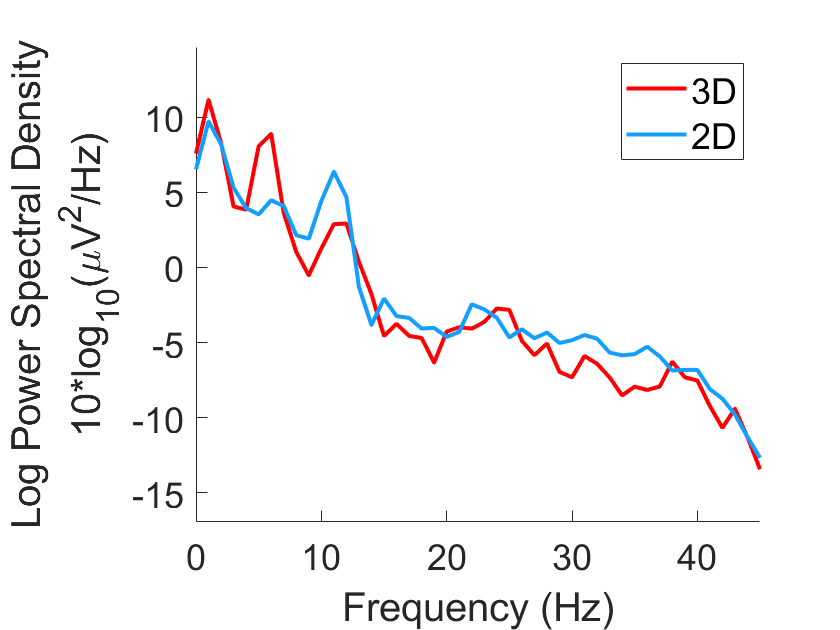}}
    \hfill
    \subfloat[Video stimuli (d)] {\includegraphics[width=0.235\textwidth]{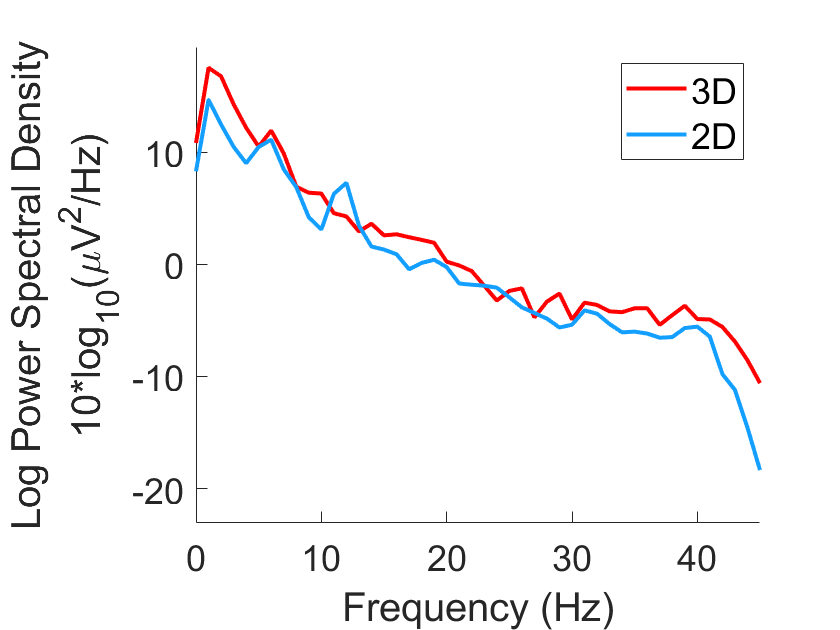}}
    \caption{The power spectral density of Pz under 2D and 3D stimuli.}
    \label{fig_pz}
\end{figure}

For the Pz electrode, the data of one subject is depicted in Fig.~\ref{fig_pz}. After statistical analysis, we derived the following statistical insights: 

\begin{itemize}

\item $Video$ $stimuli$ $(a)$: About 59\% of subjects displayed higher $\delta$ and $\theta$ oscillations power under 3D stimuli. And around 72\% of subjects showed higher $\gamma$ oscillation power under 2D stimuli.

\item $Video$ $stimuli$ $(b)$: Approximately 66\% of subjects showed higher $\alpha$ and $\beta$ oscillation power under 3D stimuli. And about 62\% of subjects showed higher $\gamma$ oscillation power under 2D stimuli.

\item $Video$ $stimuli$ $(c)$: Around 62\% of subjects experienced higher $\theta$ oscillation power under 3D stimuli.

\item $Video$ $stimuli$ $(d)$: About 69\% demonstrated higher $\alpha$ and $\beta$ oscillation power, with a similar percentage displaying higher $\gamma$ oscillation power.

\end{itemize}

The statistical results suggested that 3D video viewing might enhance $\theta$ activity, and videos involving calculation tasks led to a increased percentage of subjects showing higher $\alpha$ and $\beta$ oscillation power under 3D stimuli. Furthermore, for high-dynamic videos, the $\gamma$ oscillation power was higher under 2D stimuli.

\begin{figure}[!t]
    \centering

    \subfloat{
    \rotatebox{90}{\scriptsize{   Video stimuli (a)}}
    \begin{minipage}[t]{0.145\textwidth}
    \centering
    {\includegraphics[width=1\textwidth]{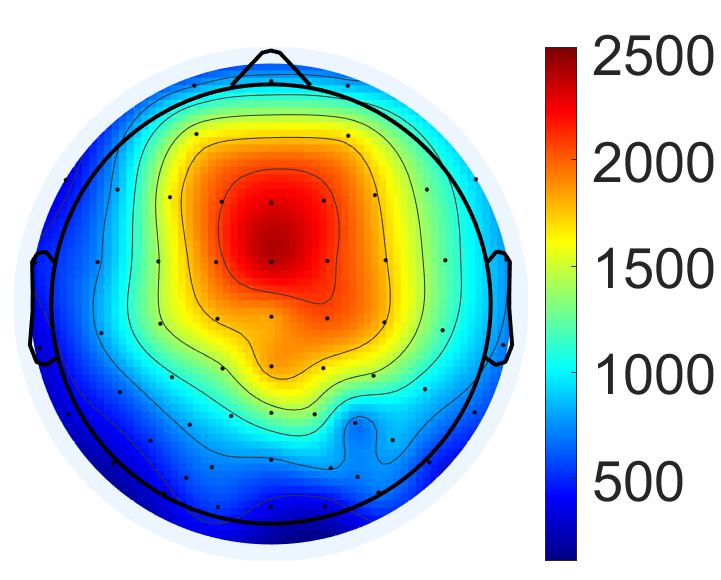}}
    \end{minipage}}
    \subfloat{
    \begin{minipage}[t]{0.145\textwidth}
    \centering
    {\includegraphics[width=1\textwidth]{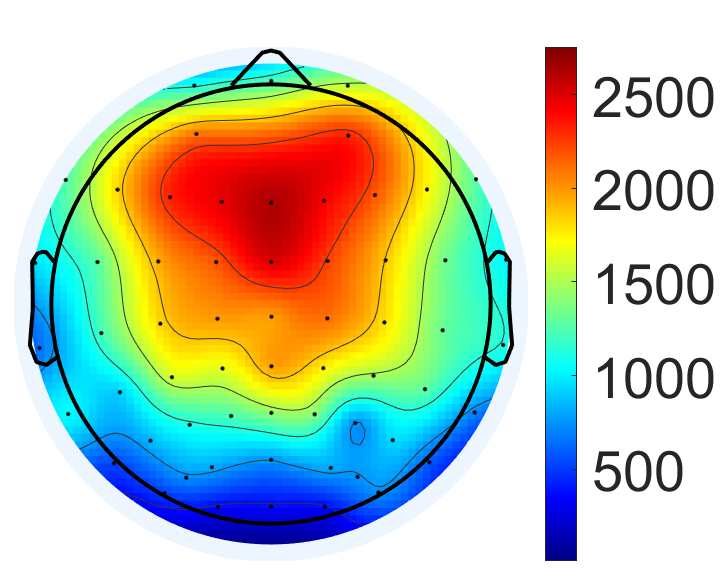}}
    \end{minipage}}
    \subfloat{
    \begin{minipage}[t]{0.145\textwidth}
    \centering
    {\includegraphics[width=1\textwidth]{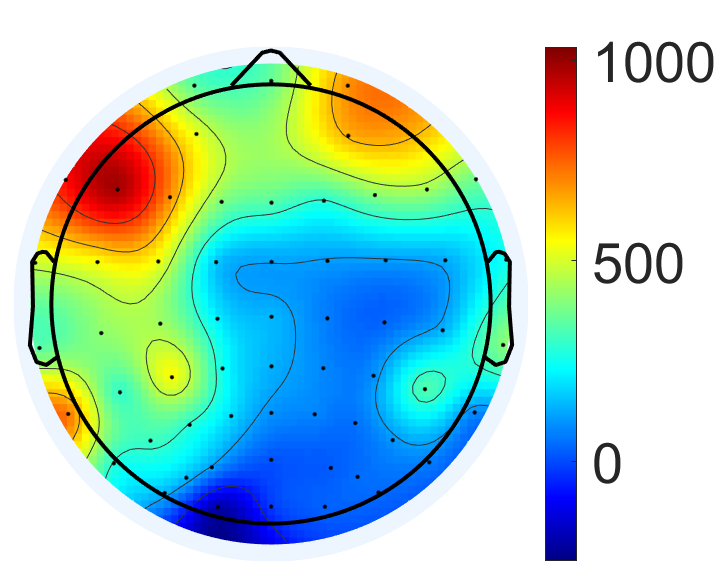}}
    \end{minipage}}

    \vspace{-3mm}

    \setcounter{subfigure}{0}

    \subfloat{
    \rotatebox{90}{\scriptsize{   Video stimuli (b)}}
    \begin{minipage}[t]{0.145\textwidth}
    \centering
    {\includegraphics[width=1\textwidth]{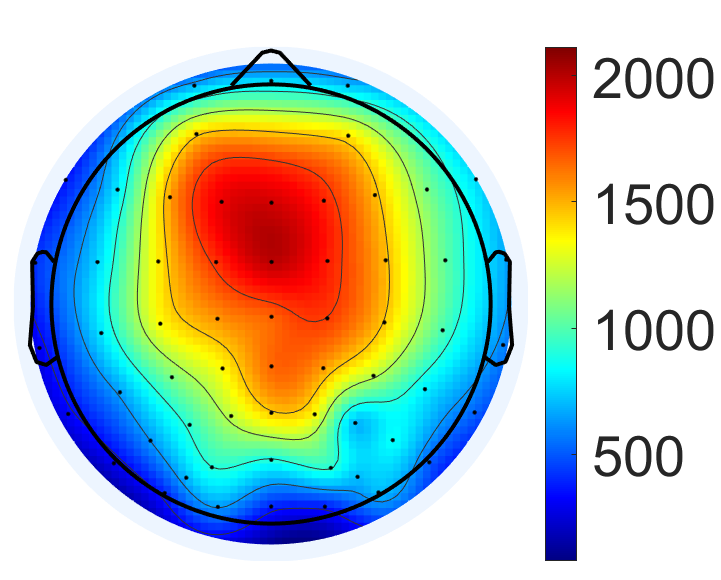}}
    \end{minipage}}
    \subfloat{
    \begin{minipage}[t]{0.145\textwidth}
    \centering
    {\includegraphics[width=1\textwidth]{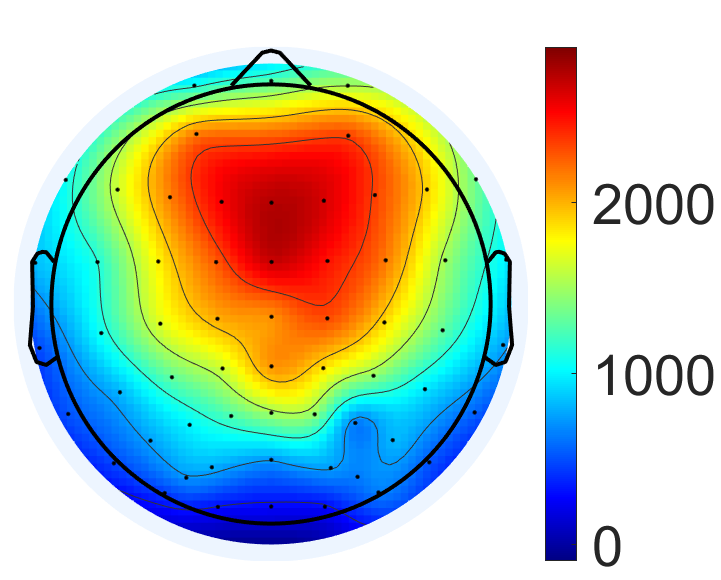}}
    \end{minipage}}
    \subfloat{
    \begin{minipage}[t]{0.145\textwidth}
    \centering
    {\includegraphics[width=1\textwidth]{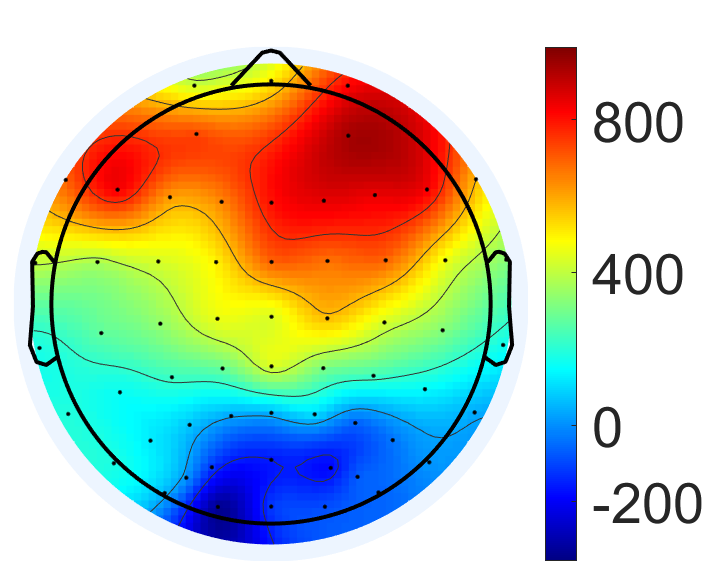}}
    \end{minipage}}

    \vspace{-3mm}

    \setcounter{subfigure}{0}

    \subfloat{
    \rotatebox{90}{\scriptsize{   Video stimuli (c)}}
    \begin{minipage}[t]{0.145\textwidth}
    \centering
    {\includegraphics[width=1\textwidth]{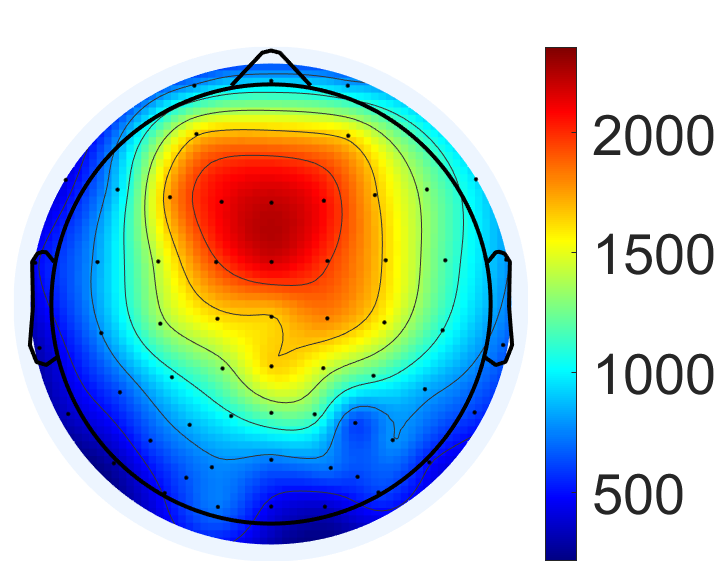}}
    \end{minipage}}
    \subfloat{
    \begin{minipage}[t]{0.145\textwidth}
    \centering
    {\includegraphics[width=1\textwidth]{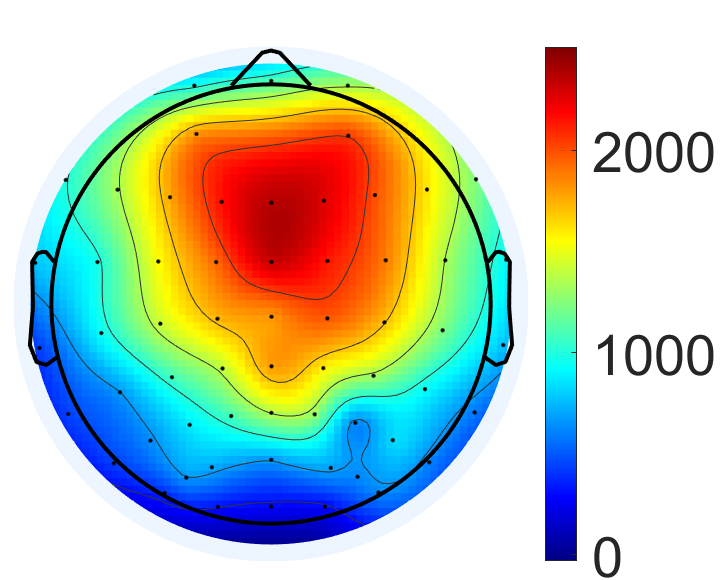}}
    \end{minipage}}
    \subfloat{
    \begin{minipage}[t]{0.145\textwidth}
    \centering
    {\includegraphics[width=1\textwidth]{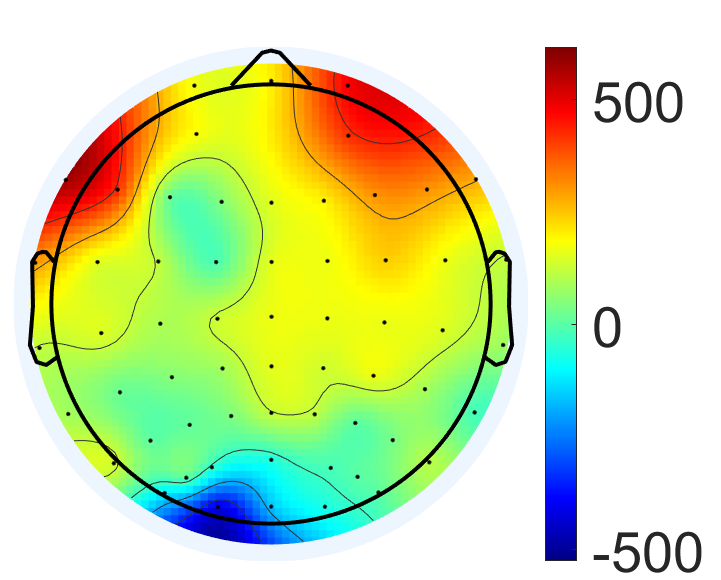}}
    \end{minipage}}

    \vspace{-3mm}

    \setcounter{subfigure}{0}

    \subfloat[2D stilmuli]{
    \rotatebox{90}{\scriptsize{   Video stimuli (d)}}
    \begin{minipage}[t]{0.145\textwidth}
    \centering
    {\includegraphics[width=1\textwidth]{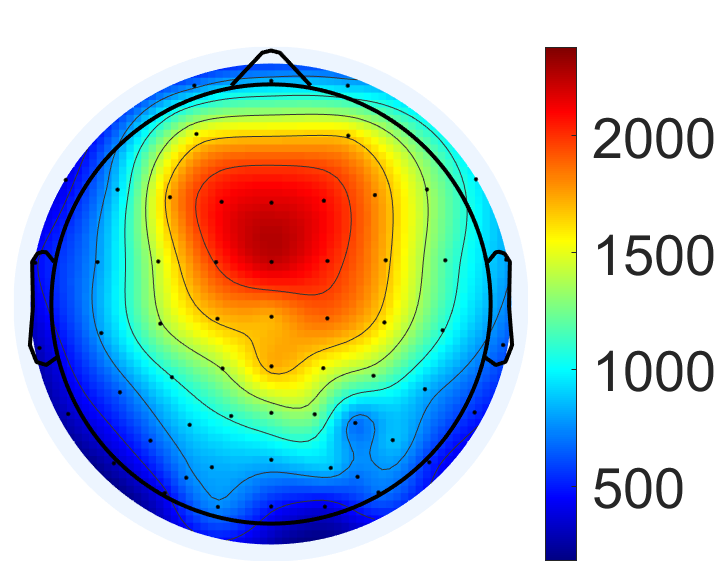}}
    \end{minipage}}
    \subfloat[3D stimuli]{
    \begin{minipage}[t]{0.145\textwidth}
    \centering
    {\includegraphics[width=1\textwidth]{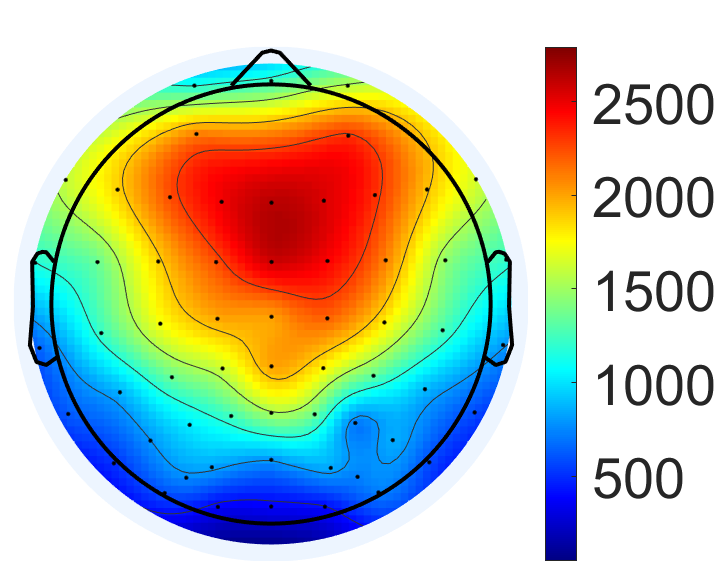}}
    \end{minipage}}
    \subfloat[D-value]{
    \begin{minipage}[t]{0.145\textwidth}
    \centering
    {\includegraphics[width=1\textwidth]{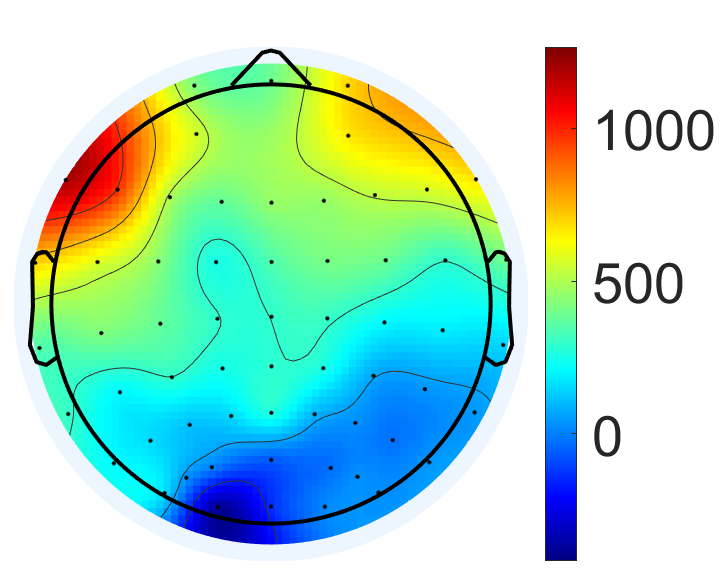}}
    \end{minipage}}

    \caption{Topographical maps and its difference under 2D and 3D stimuli.}
    \label{fig_top}
\end{figure}

Except analyzing the correlation between oscillations power at Fz and Pz electrodes and video stimuli, we also conducted a statistical analysis on the variance in brain activity across various video stimuli. Fig.~\ref{fig_top} represented the aggregate power of one subject from 0.5 to 45 Hz under 2D and 3D video stimuli, along with its difference. Our experimental results suggested that 3D videos induced greater activity in the frontal and parietal lobes compared to 2D videos. Specifically, for video (a) and (b) under 2D stimuli, activity in the parietal and central lobes was more significant than that in video stimuli (c) and (d). Likewise, an uptick in parietal lobe activity was noted in 3D videos, underscoring the association between spatial information processing and the parietal lobe. The results of video (b) and (d) under 2D stimuli demonstrated that videos with calculation tasks induced greater activity in the frontal lobe and near the Cz electrode, highlighting the cognitive role of the frontal lobe.

\subsection{$\alpha$ and $\theta$ oscillation power distribution}
\begin{figure*}[ht]
  \centering
  \subfloat[~  Video stimuli (a)]{\includegraphics[width=0.5\columnwidth]{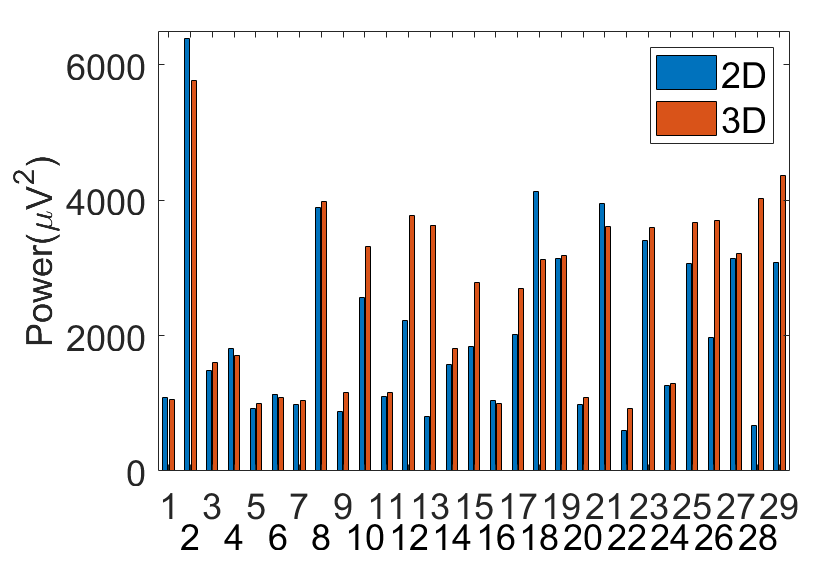}\label{fig_theta:1}}
  \hfill
  \subfloat[~  Video stimuli (b)]{\includegraphics[width=0.5\columnwidth]{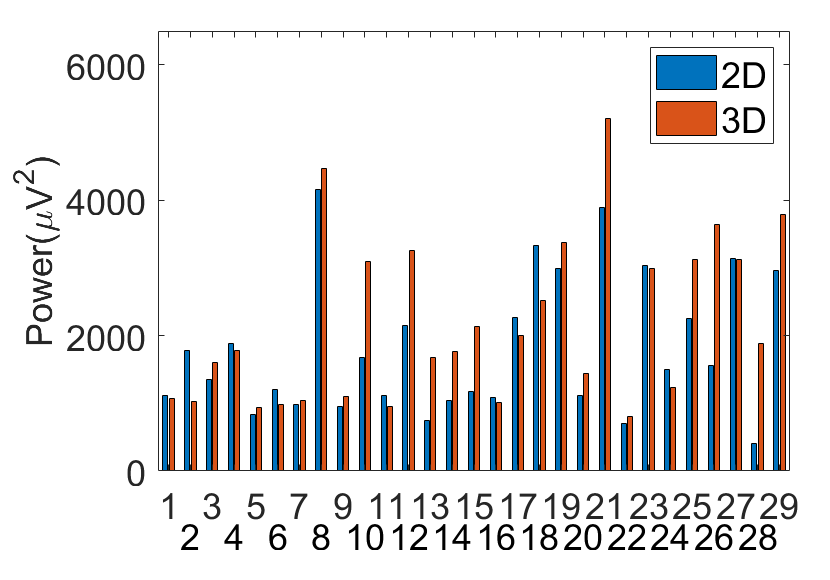}\label{fig_theta:2}}
  \hfill
  \subfloat[~  Video stimuli (c)]{\includegraphics[width=0.5\columnwidth]{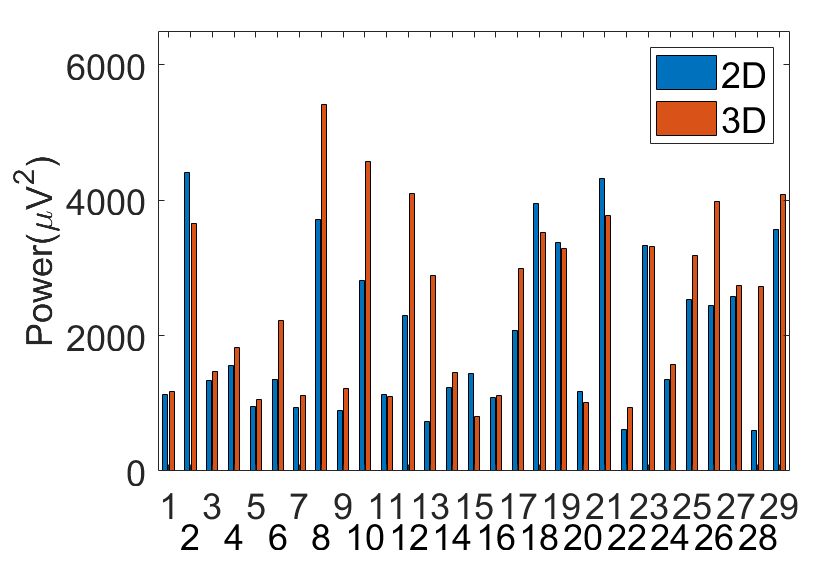}\label{fig_theta:3}}
  \hfill
  \subfloat[~  Video stimuli (d)]{\includegraphics[width=0.5\columnwidth]{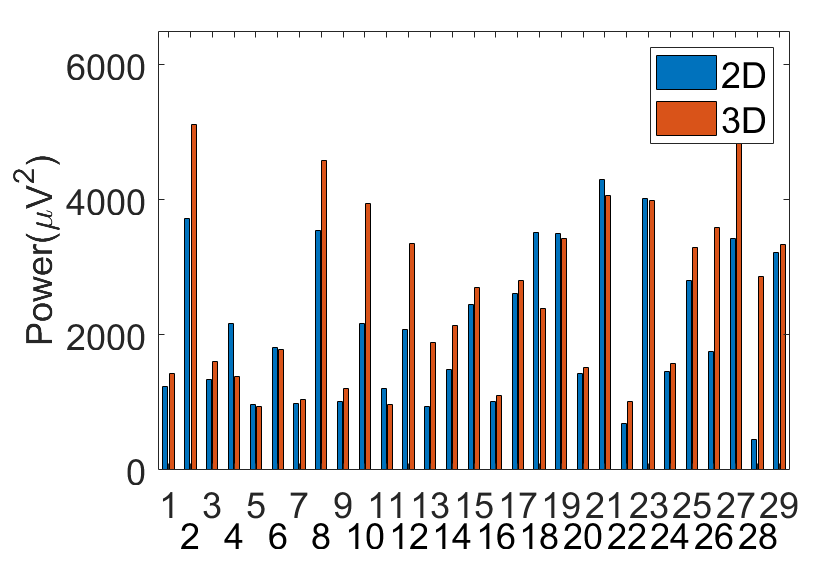}\label{fig_theta:4}}  
  \caption{
   In the figure, the x-axis represents the index of  subjects, while the y-axis represents subjects' $\theta$ oscillation power. The blue bar represents the 2D video, while the red bar represents the 3D video.}
  \label{fig_theta}
\end{figure*}

\begin{figure*}[ht]
  \centering
  \subfloat[~  Video stimuli (a)]{\includegraphics[width=0.5\columnwidth]{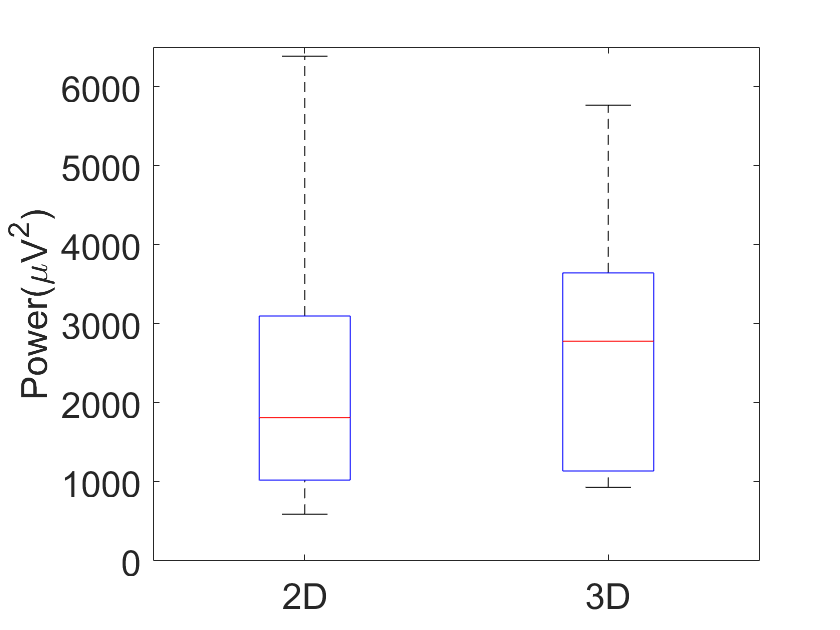}\label{fig_theta_sta:1}}
  \hfill
  \subfloat[~  Video stimuli (b)]{\includegraphics[width=0.5\columnwidth]{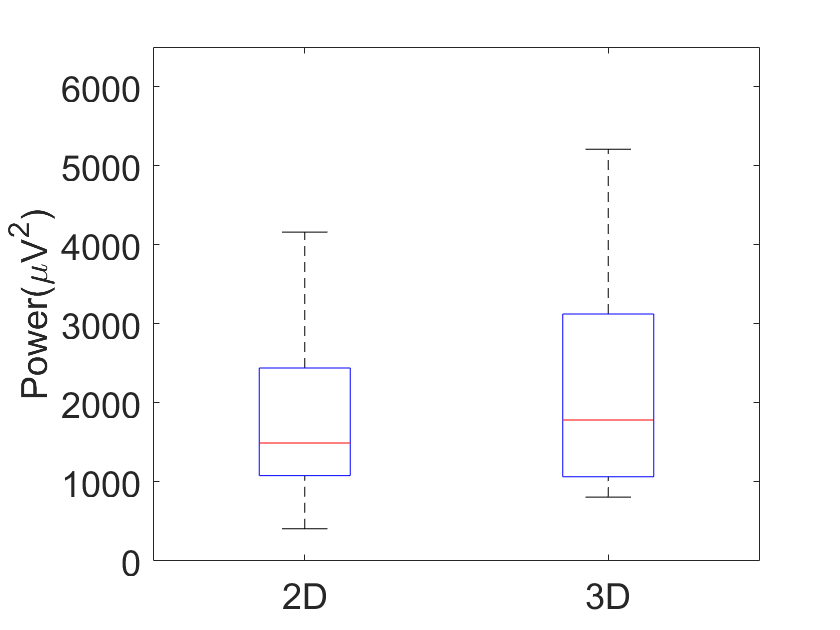}\label{fig_theta_sta:2}}
  \hfill
  \subfloat[~  Video stimuli (c)]{\includegraphics[width=0.5\columnwidth]{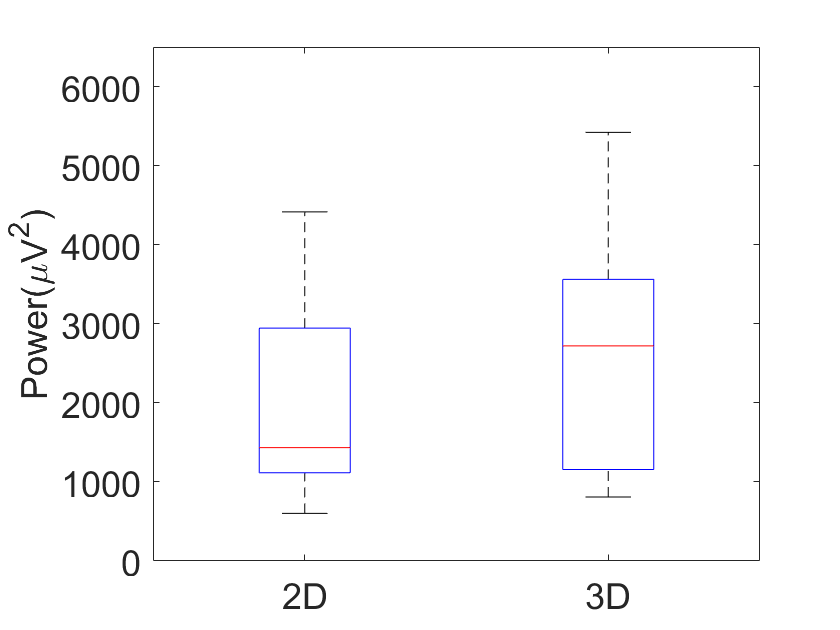}\label{fig_theta_sta:3}}
  \hfill
  \subfloat[~  Video stimuli (d)]{\includegraphics[width=0.5\columnwidth]{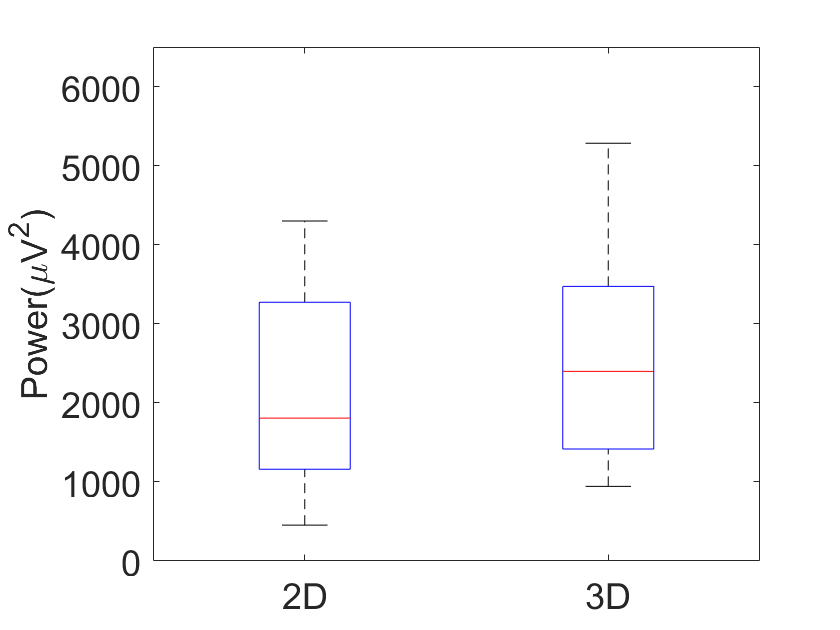}\label{fig_theta_sta:4}}  
  \caption{
  In the figure, the x-axis (2D,3D) represents 2D and 3D video stimuli, while the y-axis represents subjects' $\theta$ oscillation power.}
  \label{fig_theta_sta}
\end{figure*}

\begin{figure*}[ht]
  \centering
  \subfloat[~  Video stimuli (a)]{\includegraphics[width=0.5\columnwidth]{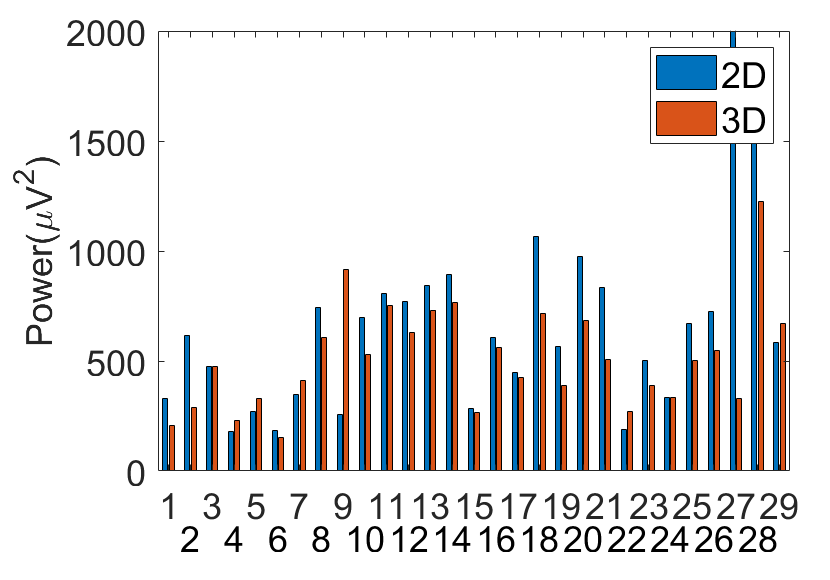}\label{fig_alpha:1}}
  \hfill
  \subfloat[~  Video stimuli (b)]{\includegraphics[width=0.5\columnwidth]{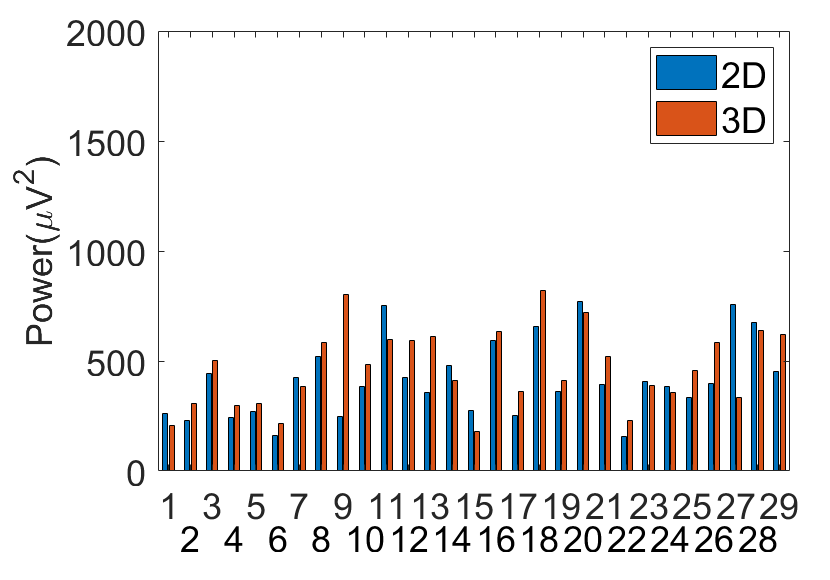}\label{fig_alpha:2}}
  \hfill
  \subfloat[~  Video stimuli (c)]{\includegraphics[width=0.5\columnwidth]{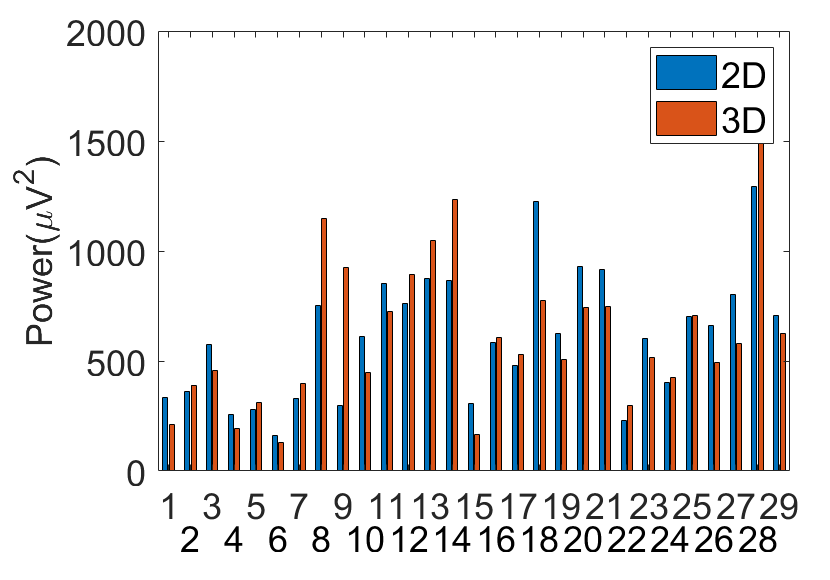}\label{fig_alpha:3}}
  \hfill
  \subfloat[~  Video stimuli (d)]{\includegraphics[width=0.5\columnwidth]{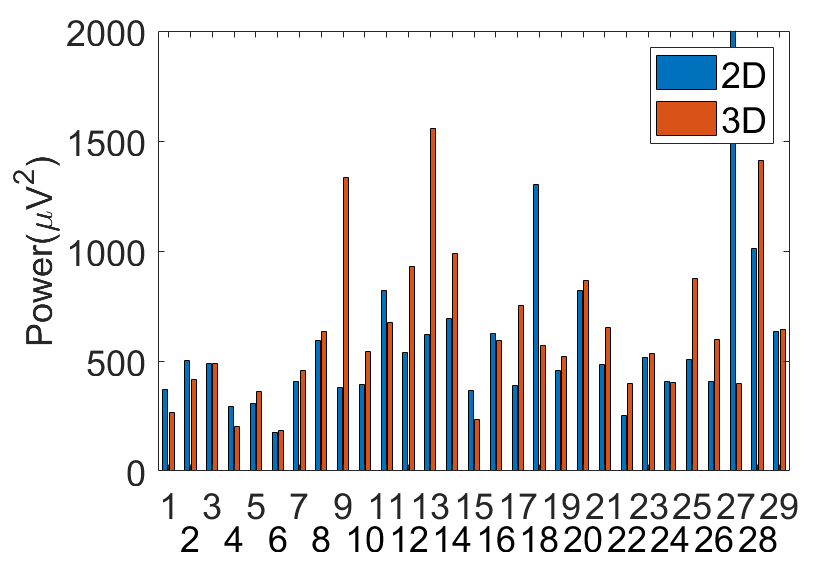}\label{fig_alpha:4}}  
  \caption{
  In the figure, the x-axis represents the index of subjects, while the y-axis represents subjects' $\alpha$ oscillation power. The blue bar represents the 2D video, while the red bar represents the 3D video.}
  \label{fig_alpha}
\end{figure*}

\begin{figure*}[ht]
  \centering
  \subfloat[~  Video stimuli (a)]{\includegraphics[width=0.5\columnwidth]{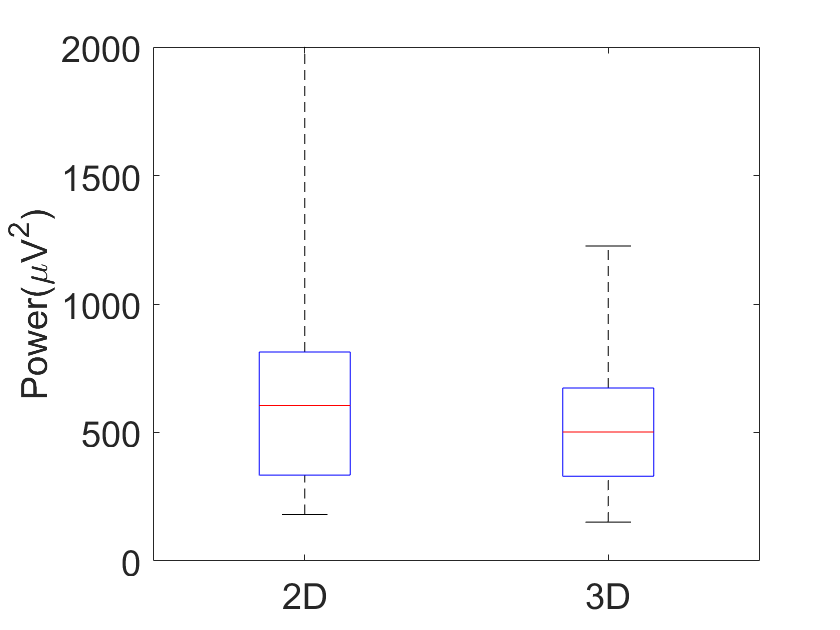}\label{fig_alpha_sta:1}}
  \hfill
  \subfloat[~  Video stimuli (b)]{\includegraphics[width=0.5\columnwidth]{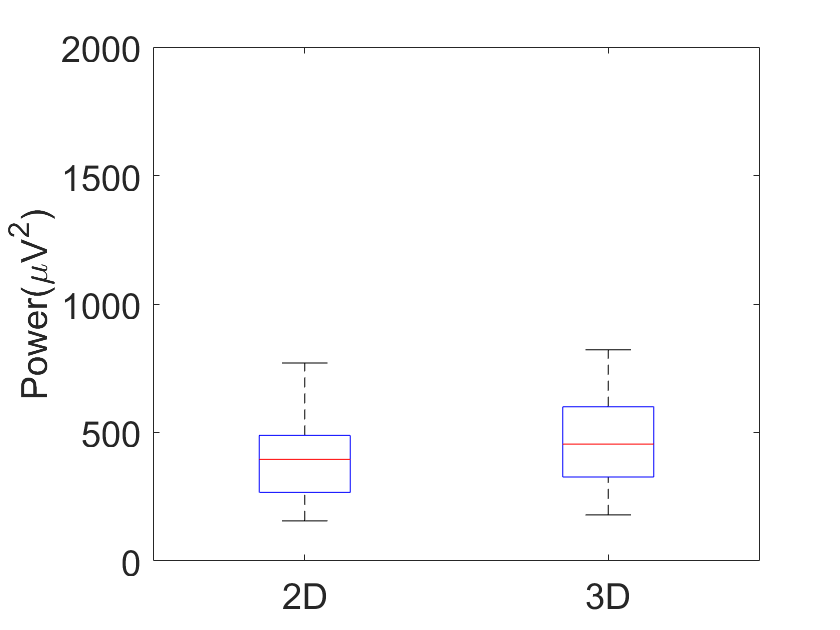}\label{fig_alpha_sta:2}}
  \hfill
  \subfloat[~  Video stimuli (c)]{\includegraphics[width=0.5\columnwidth]{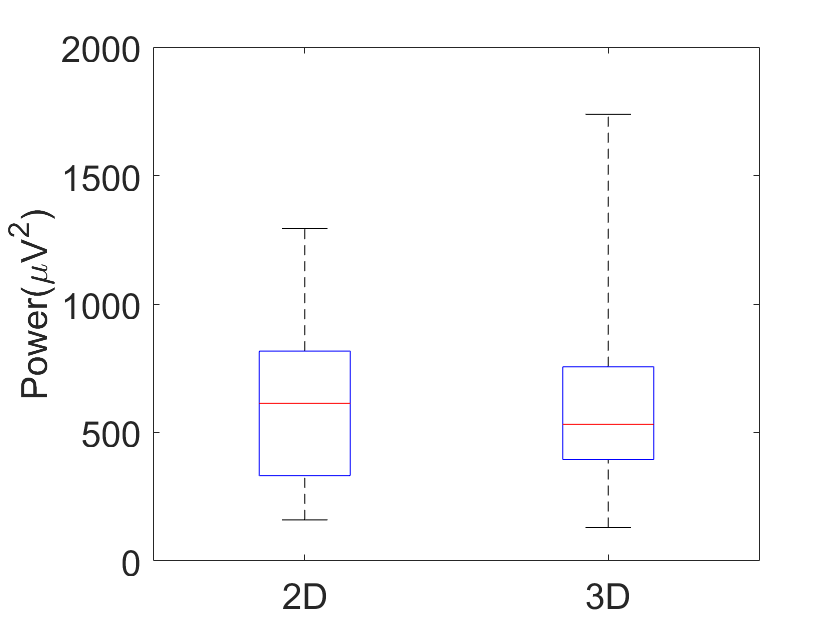}\label{fig_alpha_sta:3}}
  \hfill
  \subfloat[~  Video stimuli (d)]{\includegraphics[width=0.5\columnwidth]{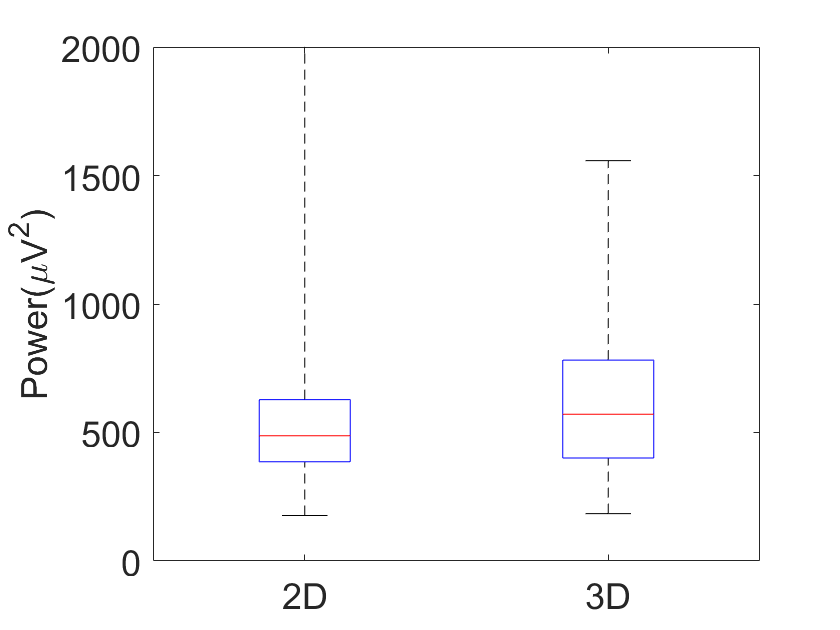}\label{fig_alpha_sta:4}}  
  \caption{
  In the figure, the x-axis (2D,3D) represents 2D and 3D video stimuli, while the y-axis represents subjects' $\alpha$ oscillation power.}
  \label{fig_alpha_sta}
\end{figure*}

Based on the EEG data collected, we computed $\alpha$ and $\theta$ oscillation power. Fig.~\ref{fig_theta} shows the $\theta$ oscillation power across the four videos under both 2D and 3D stimuli. Fig.~\ref{fig_alpha} shows the $\alpha$ oscillation power for the subjects across the same set of videos under 2D and 3D stimuli. Subsequently, we performed a statistical analysis of the experimental data. Fig.~\ref{fig_theta_sta} presents the distribution of $\theta$ oscillation power across the four videos, while Fig.~\ref{fig_alpha_sta} displays the distribution of $\alpha$ oscillation power for these videos.

Fig.~\ref{fig_theta} indicates that a majority of subjects exhibited higher $\theta$ oscillation power while viewing 3D videos, particularly notable during video stimuli (a) and (c), which were categorized as videos with simple observational tasks.

\begin{itemize}

\item $Video$ $stimuli$ $(a)$: 21\% of subjects displayed higher $\theta$ oscillation power under 2D stimuli, whereas 73\% showed higher $\theta$ oscillation power under 3D stimuli, with the $\theta$ oscillation power for the remainder being nearly identical under both 2D and 3D stimuli.

\item $Video$ $stimuli$ $(b)$: 31\% of subjects demonstrated higher $\theta$ oscillation power under 2D stimuli, while 62\% exhibited higher $\theta$ oscillation power under 3D stimuli, and the $\theta$ oscillation power for the remainder remained nearly identical under both 2D and 3D stimuli.

\item $Video$ $stimuli$ $(c)$: 21\% of subjects had higher $\theta$ oscillation power under 2D stimuli, in contrast to 69\% who had higher $\theta$ oscillation power under 3D stimuli, with the $\theta$ oscillation power for the remainder being nearly identical under both 2D and 3D stimuli.

\item $Video$ $stimuli$ $(d)$: 17\% of subjects showed higher $\theta$ oscillation power under 2D stimuli, whereas 65\% had higher $\theta$ oscillation power under 3D stimuli, and the $\theta$ power for the remainder were nearly identical under both 2D and 3D stimuli.

\end{itemize}

Fig.~\ref{fig_theta_sta} demonstrates that the mean $\theta$ oscillation power in 3D videos surpasses that in 2D videos. 
Specifically for video stimuli (a), the median $\theta$ oscillation power was more significant under 3D stimuli, whereas 2D stimuli exhibited a higher maximum value. For video stimuli (a)$-$(d), both the maximum and median $\theta$ oscillation power were higher under 3D stimuli. Disregarding the maximum value for each set, the mean $\theta$ oscillation power for these videos were slightly higher under 3D stimuli.

Fig.~\ref{fig_alpha} illustrates that the $\alpha$ oscillation power does not show a consistent pattern of difference between subjects viewing 2D and 3D videos.

\begin{itemize}

\item $Video$ $stimuli$ $(a)$: 72\% of subjects experienced higher $\alpha$ oscillation power under 2D stimuli, whereas only 21\% showed higher $\alpha$ oscillation power under 3D stimuli, with the $\alpha$ oscillation power for the remainder being nearly identical under both 2D and 3D stimuli.

\item $Video$ $stimuli$ $(b)$: 34\% of subjects demonstrated higher $\alpha$ oscillation power while viewing the 2D stimuli, compared to 65\% who showed higher $\alpha$ oscillation power under 3D stimuli, and the $\alpha$ oscillation power for the remainder were nearly identical under both 2D and 3D stimuli.

\item $Video$ $stimuli$ $(c)$: 52\% of subjects had higher $\alpha$ oscillation power under 2D stimuli, while 45\% had higher $\alpha$ oscillation power under 3D stimuli, with the $\alpha$ oscillation power for the remainder being nearly identical under both 2D and 3D stimuli.

\item $Video$ $stimuli$ $(d)$: 28\% of subjects exhibited higher $\alpha$ oscillation power under 2D stimuli, while 55\% displayed higher $\alpha$ oscillation power under 3D stimuli, and the $\alpha$ oscillation power for the remainder remained nearly identical under both 2D and 3D stimuli.

\end{itemize}

Fig.~\ref{fig_alpha_sta} reveals that the median $\alpha$ oscillation power in 2D videos surpasses that in 3D videos for video stimuli (a) and (c), whereas it is slightly higher in 3D videos for video stimuli (b) and (d).

\begin{itemize}

\item $Video$ $stimuli$ $(a)$: Both the maximum and median $\alpha$ oscillation power were higher under 2D stimuli.

\item $Video$ $stimuli$ $(b)$: The $\alpha$ oscillation power distribution was relatively even under 2D and 3D stimuli, with the maximum and median values being slightly higher under 3D stimuli.

\item $Video$ $stimuli$ $(c)$: The median $\alpha$ oscillation power was higher under 2D stimuli, but the maximum value was greater under 3D stimuli.

\item $Video$ $stimuli$ $(d)$: The median $\alpha$ oscillation power was higher under 2D stimuli, yet the maximum value was higher under 3D stimuli.
\end{itemize}

\subsection{Comparison of Cognitive Load Index}
\begin{figure*}[ht]
  \centering
  \subfloat[~  Video stimuli (a)]{\includegraphics[width=0.5\columnwidth]{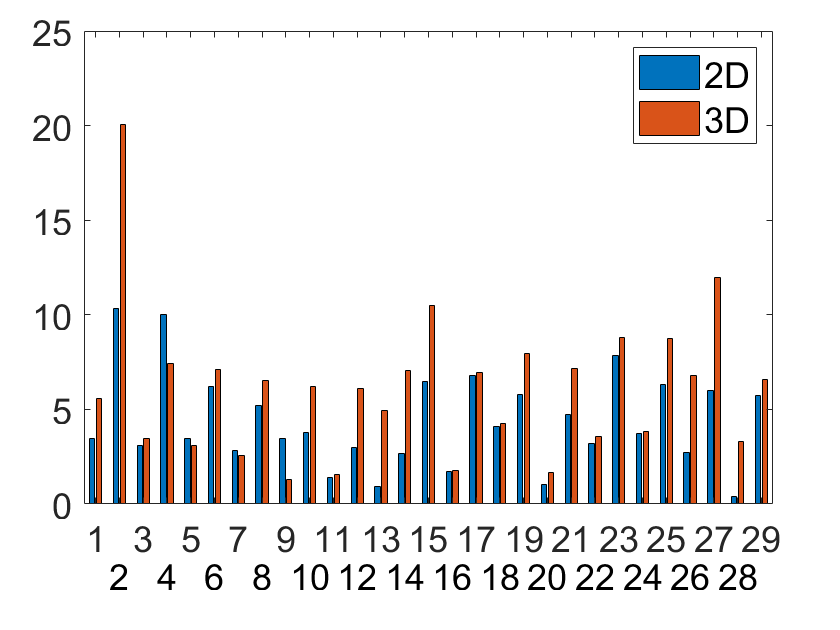}\label{fig_CLI:1}}
  \hfill
  \subfloat[~  Video stimuli (b)]{\includegraphics[width=0.5\columnwidth]{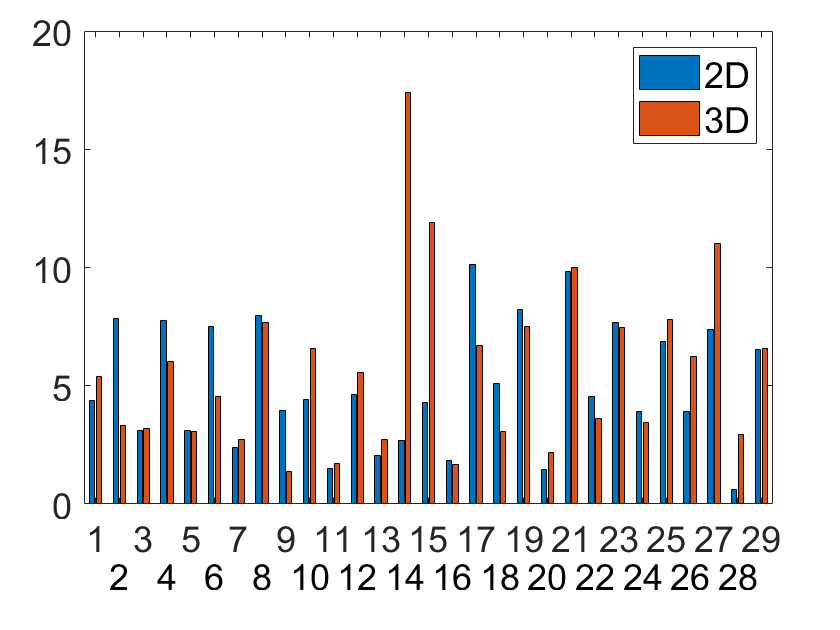}\label{fig_CLI:2}}
  \hfill
  \subfloat[~  Video stimuli (c)]{\includegraphics[width=0.5\columnwidth]{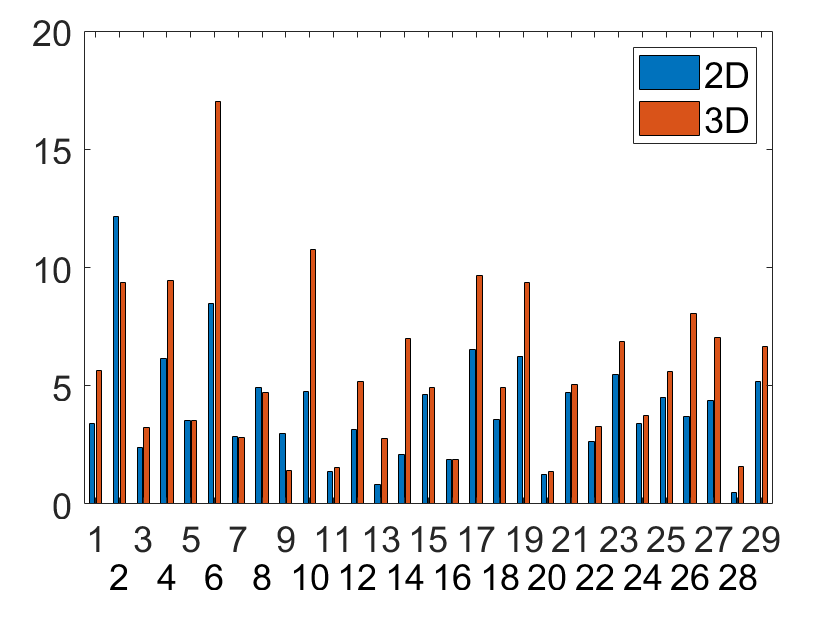}\label{fig_CLI:3}}
  \hfill
  \subfloat[~  Video stimuli (d)]{\includegraphics[width=0.5\columnwidth]{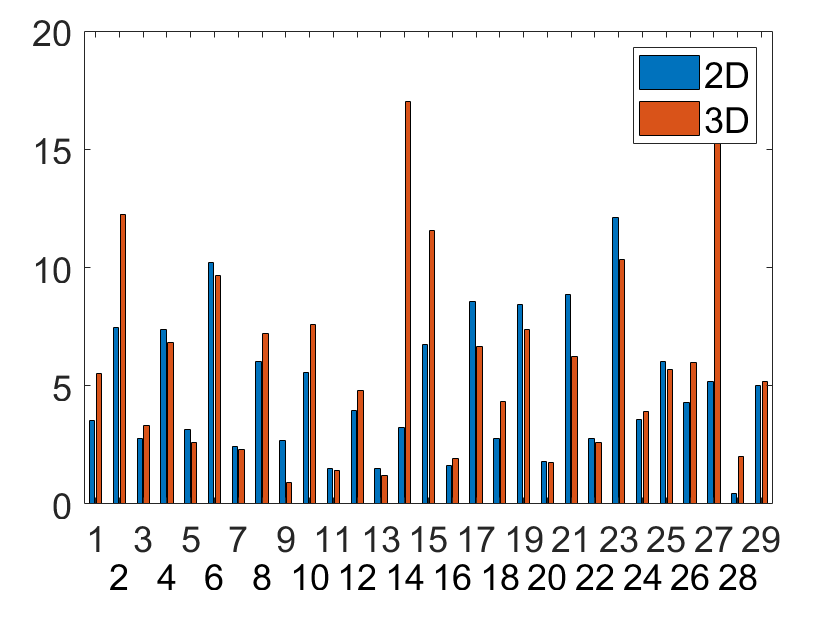}\label{fig_CLI:4}}  
  \caption{
  In the figure, the x-axis represents the index of subjects, while the y-axis represents the CLI of subjects. The blue bar represents the 2D video, while the red bar represents the 3D video.}
  \label{fig_CLI}
\end{figure*}

\begin{figure*}[ht]
  \centering
  \subfloat[~  Video stimuli (a)]{\includegraphics[width=0.5\columnwidth]{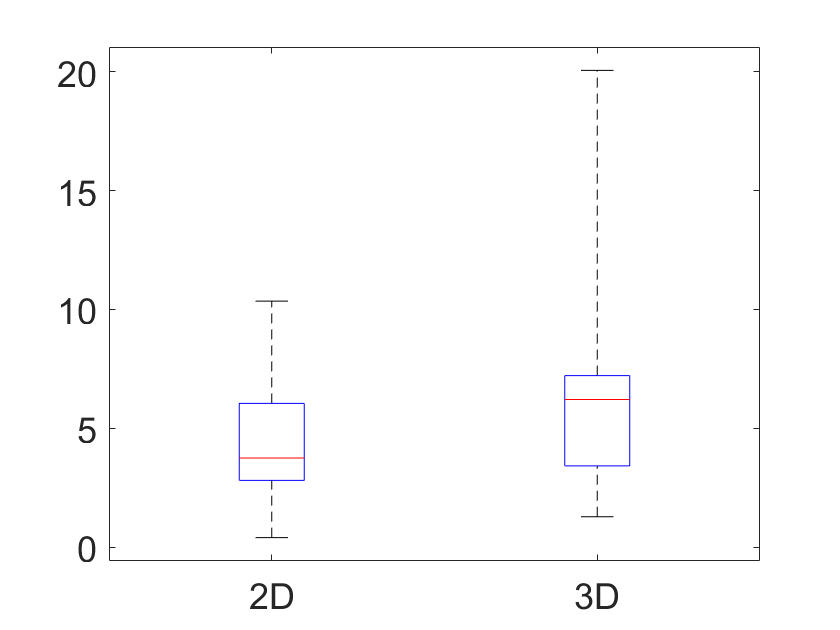}\label{fig_CLI_sta:1}}
  \hfill
  \subfloat[~  Video stimuli (b)]{\includegraphics[width=0.5\columnwidth]{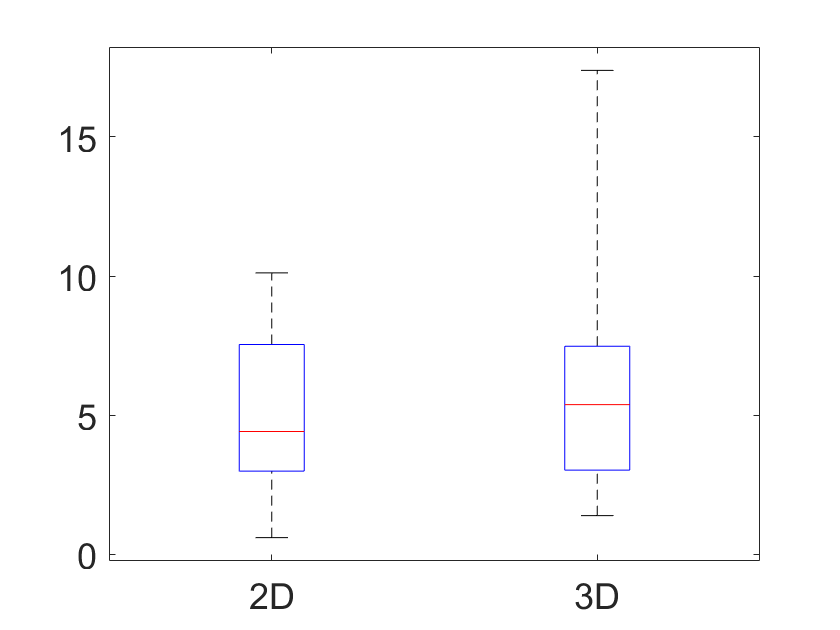}\label{fig_CLI_sta:2}}
  \hfill
  \subfloat[~  Video stimuli (c)]{\includegraphics[width=0.5\columnwidth]{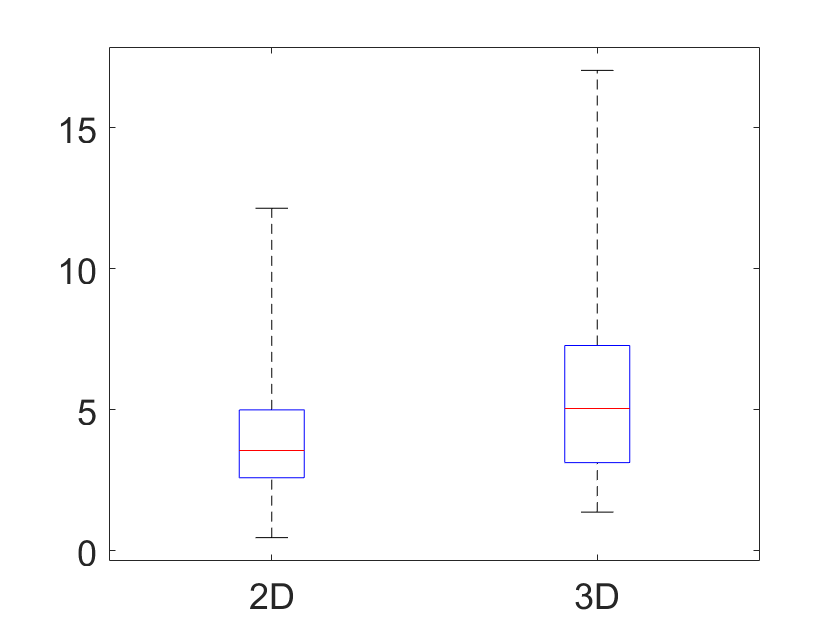}\label{fig_CLI_sta:3}}
  \hfill
  \subfloat[~  Video stimuli (d)]{\includegraphics[width=0.5\columnwidth]{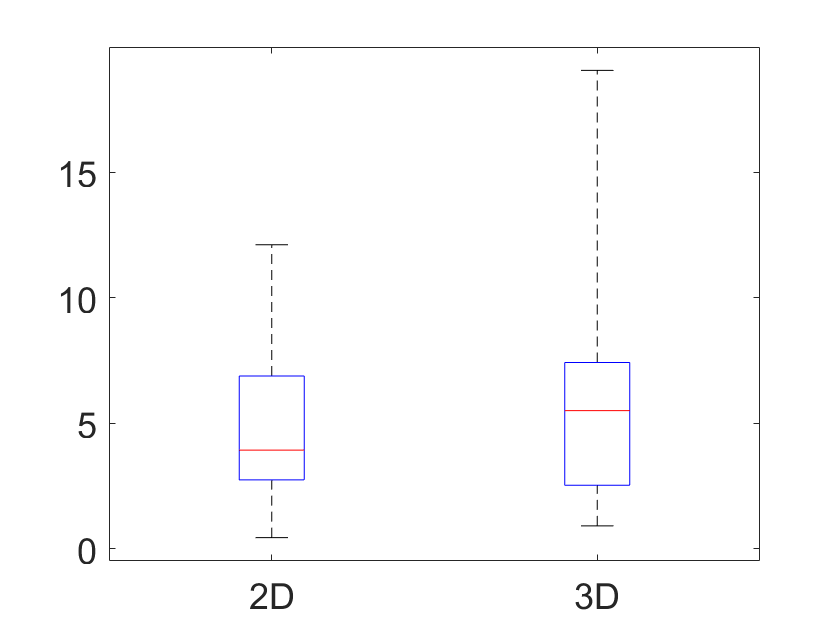}\label{fig_CLI_sta:4}}  
  \caption{
 In the figure, the x-axis (2D,3D) represents 2D and 3D video stimuli, while the y-axis represents the CLI of subjects.}
  \label{fig_CLI_sta}
\end{figure*}

The CLI statistic results were shown in Fig.~\ref{fig_CLI}.
The results revealed that CLI characteristics were highly similar to those of $\theta$ oscillation power. 
Specifically, for video stimuli (a) and (c), approximately 72\% of subjects exhibited a higher CLI while viewing the 3D stimuli.
While the exact numbers varied, the pattern of CLI variation when viewing 2D and 3D stimuli of video (a) and (c) remained nearly identical; that is, subjects who exhibit a higher CLI under 3D stimuli of video (a) also tended to have a higher CLI under 3D stimuli of video (c). Whereas, no significant statistical differences were observed for video stimuli (b) and (d). 
Furthermore, CLI values across subjects showed notable individual variability. The mean and variance of the CLI for each video were presented in Table \ref{tab_t_test}.
According to the table, all four videos demonstrated a higher mean CLI under 3D stimuli, with the difference between 2D and 3D stimuli being more significant for video stimuli (a) and (c).

Fig.~\ref{fig_CLI_sta} illustrates the distribution of CLI when viewing the four videos under 2D and 3D stimuli.
From these observations, it was apparent that the maximum and median CLI when viewing the 3D stimuli of all four videos were comparatively higher. Moreover, the distribution of CLI for these video stimuli tended to be more concentrated under 2D stimuli.
To further verify the relationship between CLI and videos, we performed an unpaired T-test analysis with $P$-value \textless 0.05 to test the significance of the difference between 2D and 3D stimuli. The statistical results were displayed in Table \ref{tab_t_test}. The results indicated that there was a significant difference in CLI between 2D and 3D for video stimuli (a) ($P$ = 0.045) and video stimuli (c) ($P$ = 0.047), whereas the differences for the video stimuli (b) ($P$ = 0.460) and video stimuli (d) ($P$ = 0.198) were not statistically significant.

\begin{table*}[!t]
    \caption{Analysis of the differences between CLI in four videos under 2D and 3D stimuli}
    \centering
    \tabcolsep=10pt
    \begin{tabular}{ccccccc}
        \toprule
        \textbf{Dependent Variable} & \textbf{Argument}& \textbf{Sample Size}& \textbf{mean Value}& \textbf{Standard Deviation}& \textbf{$t$}& \textbf{$p$}\\
        \midrule
        \multirow{2}{*}{CLI (a)} & 2D & 29 & 4.3579 & 1.5815 & \multirow{2}{*}{-2.049} & \multirow{2}{*}{\bm{$0.045$}}\\
         & 3D & 29 & 6.0947 & 3.8183 &  & \\
        \midrule
        \multirow{2}{*}{CLI (b)} & 2D & 29 & 5.0204 & 1.6269 & \multirow{2}{*}{-0.743} & \multirow{2}{*}{0.460}\\
         & 3D & 29 & 5.6378 & 1.899 &  & \\
         \midrule
        \multirow{2}{*}{CLI (c)} & 2D & 29 & 4.0567 & 1.5525 & \multirow{2}{*}{-2.031} & \multirow{2}{*}{\bm{$0.047$}}\\
         & 3D & 29 & 5.6683 & 1.8786 &  & \\
         \midrule
        \multirow{2}{*}{CLI (d)} & 2D & 29 & 4.8122 & 1.7091 & \multirow{2}{*}{-1.303} & \multirow{2}{*}{0.198}\\
         & 3D & 29 & 6.1161 & 4.5272 &  & \\
        \bottomrule
    \end{tabular}
    \begin{tablenotes}
        \item[1] The CLI (a)-(d) in the table corresponds video stimuli (a)$-$(d), respectively.
    \end{tablenotes}
\label{tab_t_test}
\end{table*}

\section{Discussion}
\label{sect_Discussion}
This paper aimed to investigate cognitive differences between subjects viewing 2D and 3D videos across various video stimuli.
Despite the 3D video complies better with human visual system, Viewing 3D movies is more apt to cause discomfort and fatigue compared to 2D counterparts. It is suggested that the additional depth cues in 3D videos require more mental resources to process the visual information.
Our results indicated that for videos involving simple observational tasks, the cognitive load was higher for subjects viewing 3D videos compared to 2D counterparts. This suggested that viewing 3D videos necessitated the allocation of more mental resources for the basic tasks of perception and understanding of the content. However, no difference was observed in situations where subjects were engaged in calculation tasks while viewing the videos.
Interestingly, the intensity of variations in video content did not appear to affect the cognitive load differences between subjects under 2D and 3D video stimuli, in contrast to the significant effect of task complexity.

All experiments were conducted utilizing a 2D/3D playback device. Distinguishing itself from other 3D video playback devices, our device can achieve 3D playback with an obstructed view (meaning only video stimuli are visible within the user's field of view), ensuring that the collected EEG signals stem exclusively from the video stimuli provided by the experiment. To verify the presence of differences in the EEG signals corresponding to various video stimuli, we analyzed the power distribution across various oscillations for the EEG signals associated with the four videos. The statistical results of Fz and Pz electrodes suggested discernible differences in EEG signals between 2D and 3D videos.
Based on the analysis of the Fz electrode, the following conclusions have been drawn:

\begin{itemize}
\item 3D videos can induce more $\theta$ and $\alpha$ activity.

\item Performing calculation tasks during video stimuli can induce more $\beta$ activity.

\item Tranquil videos can induce more $\delta$ activity.

\end{itemize}

Regarding the Pz electrode, the following conclusions have been drawn:

\begin{itemize}
\item 3D videos can induce more $\theta$ activity.

\item Performing calculation tasks during video stimuli can induce more $\alpha$ activity.

\item High-dynamic videos can induce more $\gamma$ activity.

\end{itemize}

According to the experimental results, some subjects exhibited substantially higher $\theta$ oscillation power compared to others, a trend also observed in $\alpha$ oscillation power. Despite their $\theta$ oscillation power being significantly higher than that of others, its value still remained within the normal range. From the perspective of video task, the analysis indicated that calculation-intensive tasks generally required more mental resources for cognitive processing, whereas tranquil videos tended to diminish these demands.
It was evident that we could not conclude that everyone experienced a higher cognitive load while viewing 3D videos. Notably, even with the video stimuli (a) and (c), a subset of subjects exhibited a higher cognitive load while viewing the 2D stimuli. Furthermore, the CLI among subjects showed noticeable individual variability, with the standard deviation of CLI exceeding 38\% of the mean value.

Kalyuga et al. suggested that cognitive load, when matched with the learners' capabilities and requirements, could lead to the efficient execution of learning tasks \cite{kalyuga2009instructional}.
Leveraging the insights gained from this study, it seemed conceivable that for brief educational sessions or experiments with simple video stimuli, utilizing 3D videos could effectively intensify cognitive load, thereby enhancing task performance.
This proposition hold particular relevance in experiments investigating cognitive states through eye movement, where video stimuli duration is brief. Additionally, given that tasks in such experiments primarily entail simple observations without calculation elements, the use of 3D video stimuli can optimally mobilize brain resources, thus offering a more profound insight into the cognitive states of subjects.

\section{Conclusion}
\label{sect_conclusion}

In this study, we utilized a 2D/3D playback device to limit the viewer's field of vision while presenting video stimuli, with the goal of investigating the cognitive load differences between 2D and 3D stimuli across various video stimuli. The Cognitive Load Index (CLI) was used as the metric to assess the cognitive load induced by 2D and 3D stimuli. Compared to 2D stimuli, 3D stimuli showed an increase in $\alpha$ and $\theta$ oscillation activity in the frontal lobe, as well as enhanced $\theta$ oscillation activity in the parietal lobe. The experimental results suggested that more dynamic video content tends to induce greater brain activity. T-test comparisons of the CLI for 2D and 3D stimuli revealed that tasks requiring simple observation resulted in a significantly higher cognitive load in subjects viewing 3D stimuli. Thus, we believe that although the cognitive load may seem similar under 2D and 3D stimuli, 3D videos associated with observation tasks generally induce a higher cognitive load.

\section*{Acknowledgments}
This work is partly funded by the National Natural Science Foundation of China.



\bibliographystyle{IEEEtran}
\bibliography{Ref}

\vfill

\end{document}